\documentclass[12pt,a4paper]{article}
\usepackage[top=1.5in, bottom=1.4in, left=0.93in, right=0.87in]{geometry}
\usepackage{amsmath,amssymb,amsfonts}
\usepackage[colorlinks]{hyperref}
\usepackage{graphicx}
\usepackage{array}
\usepackage{feynmf,bookmark}
\usepackage{ulem}

\newcommand{\eq}[1]{Eq.~\eqref{#1}}
\newcommand{\Figref}[1]{Fig.~\ref{#1}}
\newcommand{\Tabref}[1]{Tab.~\ref{#1}}
\newcommand{\Secref}[1]{Sec.~\ref{#1}}

\def\bea{\begin{eqnarray} }
\def\eea{ \end{eqnarray} } 

\def\be{\begin{equation} }
\def\ee{ \end{equation} }

\newcommand{ \Cpm       }{\tilde{\chi}^\pm}

\newcommand{\BRBsm}{\text{BR}(B_s\rightarrow\mu^+\mu^-)}
\newcommand{\BRBdm}{\text{BR}(B_d\rightarrow\mu^+\mu^-)}

\newcommand{\Bsuntg}{\overline{\text{BR}}(B_s\rightarrow\mu^+\mu^-)}
\newcommand{\sflav}{\rm{SUSY\_FLAVOR}}

\newcommand{\ADF} {{\mathcal{A}}^{\mu\mu}_{\Delta\Gamma}}
\newcommand{\One}{{\mathbf{1}}} 

\newcommand{\gev}{\:\text{GeV}}
\newcommand{\tev}{\:\text{TeV}}

\usepackage{cleveref}
\crefformat{section}{\S#2#1#3} 
\crefformat{subsection}{\S#2#1#3}
\crefformat{subsubsection}{\S#2#1#3}

\newcommand{\Bsm}{B_s\rightarrow\mu^+\mu^-}

\begin{document}

\begin{titlepage}
\vspace{2cm}
\begin{center}
  {\Large\bf{30 parameter version of the MSSM in light of the
      $B_s\rightarrow\mu^+\mu^-$ observation 
  }}

\vspace{1cm}
\renewcommand{\thefootnote}{\arabic{footnote}}

\textbf{
S.S. AbdusSalam\footnote[1]{Email:
  \texttt{abdussalam@sbu.ac.ir}}$^{(a)}$ and 
L. Velasco-Sevilla\footnote[2]{Email: 
\texttt{Liliana.Velasco-Sevilla@ift.uib.no}}$^{(b)}$
}
\\[5mm]
\end{center}
\textit{\small
{$^{(a)}$ Department of Physics, Shahid Beheshti University, Tehran 19839, Islamic Republic of Iran\\[2mm]
$^{(b)}$ University of Bergen, Department of Physics and Technology, POB 7803, 5020 Bergen, Norway\\[2mm]
}}

\vspace{1cm}

\begin{abstract}
  The R-parity conserving MSSM in light of the decay $B_s\rightarrow\mu^+\mu^-$ with near-SM branching ratio is an interesting platform for studying the complementarity between direct and indirect searches for beyond the SM physics. Based on this, we have analysed the possible impact of the $B_s\rightarrow\mu^+\mu^-$ observation on the posterior sample from the global fit of a 30-parameter MSSM (MSSM-30), and the related Wilson   Coefficients. The MSSM-30 is a systematically constructed, symmetry-guided, MSSM parametrization, as opposed to the traditional frames (e.g. pMSSM) with crude treatment of flavor violation parameters. This paper illustrates why phenomenological frames like the MSSM-30 should be preferred to study flavor physics. For the current and future B-physics experimental precision, such a consideration is crucial for suitably assessing supersymmetric contributions to flavor observables.
\end{abstract}

\end{titlepage}

\tableofcontents

\section{Introduction}

The decay $\Bsm$ has been used traditionally as an
indicator of how contributions from extended 
Higgs sectors, with respect to the Standard Model (SM), can give sizeable
contributions to leptonic decays. This happens because its branching fraction undergoes  
a helicity suppression by
$m^2_{\mu}/M^2_{B_s}$, where $m_\mu$ is the mass of the muon and
$M_{B_s}$ is the mass of the $B_s$ meson. 
This helicity suppression can be
lifted in models with extra Higgs doublets where chirality-changing quark flavor violation is present and it is particularly strong for large values of $\tan\beta$  \cite{Bobeth:2001sq,Bobeth:2002ch}. The chirality-changing quark flavor violation contributions to $\Bsm$  are proportional to $m^2_{B_s}/m^2_b$ instead.  Hence, these decays provide a good opportunity to look for physics beyond the Standard
Model (BSM). For the particular BSM case of 
 the R-parity conserving minimal supersymmetric standard
model (MSSM), with diagonal soft-squared mass matrices and trilinear terms, analytical approximations indicate that the $\Bsm$ decay
amplitude can be proportional to $\tan^3 \beta$ \cite{Huang:1998vb,
  Hamzaoui:1998nu, Choudhury:1998ze, Babu:1999hn, Huang:2000sm}.
Here $\tan \beta$ is the ratio of the MSSM Higgs fields vacuum
expectation values:  $\left<H_2\right>/\left<H_1\right>$ that can take
values 
between 2 and 60. However, full-fledged (numerical) analyses including
global fits of models to experimental data have shown that  BSM contributions to
$\Bsm$ behave in a  multi-dimensional manner.

Experimental  constraints mostly suppress, or require opposite signs with
similar magnitudes, the various supersymmetric contributions with
respect to the SM one. Examples showing the manifestation of this suppression 
 were shown within the phenomenological MSSM (pMSSM) framework 
\cite{AbdusSalam:2009qd, Arbey:2012ax}, where
the $\BRBsm$ posterior distribution lies around the SM
prediction despite the moderately high values of $\tan \beta$, and in
the more recent work of \cite{Altmannshofer:2017wqy},
as required by experiments which indicate the absence of
  large deviations of $\BRBsm$ from the SM prediction
  \cite{Aaltonen:2011fi, Abazov:2010fs, Aaij:2012ac,
    Chatrchyan:2012rga, Aad:2012pn, Aaij:2012nna, CMS:2014xfa,
    Mulder:2017hug}.

Given the amazing consistency of the predictions of the SM with flavor
observables one may wonder if in supersymmetry there is a mechanism, that
just as it happens in the SM,  effectively forbids flavor changing neutral current
 (FCNC) processes and controls CP violation. Specific models with a MSSM spectrum which avoid FCNC and CP
violating processes can be constructed successfully
\cite{Kadota:2011cr,Ellis:2016qra,Poh:2015wta}.  However, 
 without a specific model for generation of flavor within the MSSM, 
 it is necessary to work within a phenomenological framework with a systematically constructed
parametrization of flavor violation.  Following this rationale, we
consider a MSSM framework with 30 parameters \cite{Colangelo:2008qp,
  AbdusSalam:2014uea} which goes
    beyond the constrained MSSM (see
\cite{Dedes:2001fv, Ellis:2005sc, Heinemeyer:2008fb, Alok:2009wk,
  Mahmoudi:2012un, Buchmueller:2012hv, Workgroup:2017myk} for related works
within the constrained MSSM set ups)
and the pMSSM \cite{Djouadi:1998di, AbdusSalam:2008uv,
  AbdusSalam:2009qd} where  flavor violation in the SUSY breaking mass terms
 is  manipulated by hand albeit with reasonable motivations.

 In the pMSSM there is no information about how flavor violation in the soft-squared masses and trilinear terms generated by radiative corrections will impact flavor observables. We know that once full 3 $\times$ 3 Yukawa matrices are considered, flavor violation is automatically generated through radiative effects. In the MSSM-30,  
a counting rule  keeps track of the hierarchical structure of the Yukawa
matrices, which are expanded in terms of the Cabibbo angle. Then trilinear terms and soft squared masses can be
expanded in that basis.  The off-diagonal parameters generated in that way can be thought of as the effective off-diagonal parameters generated through radiative corrections. This rationale discards  dangerous terms for
FCNC and CP violation. 
 In this sense, this work builds further on the project for 
MSSM explorations within systematically built frames, in this case a specific frame for flavor violation,
deriving inference from experimental data
 \cite{AbdusSalam:2008uv, AbdusSalam:2009qd,
  AbdusSalam:2010qp, AbdusSalam:2011hd, AbdusSalam:2011fc,
  AbdusSalam:2012sy, AbdusSalam:2012ir, AbdusSalam:2013qba,
  AbdusSalam:2014uea, AbdusSalam:2015uba}  with fewer theoretical or traditional prejudices
compared to 
other MSSM phenomenology frameworks. 

The structure of flavor violation{\footnote{The way flavor is set up at  in \cite{Colangelo:2008qp}  takes
$Y_u=\lambda_u V $ and $Y_d$  diagonal, where $\lambda_u$ is the
diagonal matrix of Yukawa couplings and $V$ is the
Cabibbo-Kobayashi-Maskawa (CKM) matrix.}} and the flat distribution chosen for this work  make our results obviously model dependent but one of our points is to exemplify how
a realistic treatment for flavor effects in the MSSM can influence the allowed parameter space regions of supersymmetric parameters.
The flavor structure that we have chosen is a realisation of minimal flavour violation (MFV), and within this framework the posterior sample from the MSSM-30 fit
in \cite{AbdusSalam:2014uea} indicates results which are more restrictive than 
  those from the LHC searches for gluinos and squarks. 
  The MFV parametrization favours heavy gluinos and squarks
  \cite{AbdusSalam:2014uea}, in order to satisfy flavor and electric
  dipole moment constraints.

 This work is organised as follows. In \Secref{OnmssmParams}, 
  we present a
 brief review of  the MSSM-30 construction and the global fit of its
 parameters to data from indirect searches for BSM physics. The
   effect of the $\Bsm$ measurement, and the possible future accuracy of
   the measurement, on the MSSM-30 parameters is analysed there. In
 \Secref{sec:Bsmupmum}, we present  
respectively the numerical anatomy and analyses of the Wilson
Coefficients, in terms of contributions to $\Bsm$ classified according to kind of diagrams and kind of particles.
For some interesting points, we give comparisons and contrast the 
MSSM-30 results to the pMSSM case. The summary and conclusions are
presented in \Secref{sec:summary}.

\section{The MSSM-30  parameters in light of the $\Bsm$ observation}
\label{OnmssmParams}
Here we briefly set the context of our
analyses. First, the 30-parameter-MSSM framework is presented and contrasted with 
the pMSSM 
giving emphasis to the constraints on the two parameters
 most sensitive to the $\BRBsm$ observable: $\tan\beta$ and $m_A$. Second, the
possible impact of future experimental precision in the $\BRBsm$
measurement is addressed. {{Finally, we comment on the
possible impact of a future $\BRBsm$ precision measurement on the MSSM
$(m_A, \tan \beta)$ plane.}}  

\subsection{The MSSM-30 frame}
In \cite{AbdusSalam:2014uea} the 
minimal flavor violation MSSM parameters selection 
scheme leads to an MSSM frame with 30 parameters:
\bea
\label{30parameters}
\underline{\theta} \equiv \{ \, & &\!\!
{\rm{Re}} \left[{\widetilde{M}}_{1,2}\right]\!\!,  \ \ M_3, \,\quad M_A, \,\quad \tan \beta, \,\quad
{\rm{Im}}\left[\widetilde{M}_{1,2}, \,\quad \tilde{\mu}\right], \quad {a}_{1,2,3,6,7}, \,\quad {\rm{Re}}\left[\tilde{a}_{4,5,8}\right], \\\nonumber
& & {\rm{Im}}[\tilde{a}_{4,5,8}], \,\quad x_{1,2}, \, \quad y_{1,3,6,7}, \quad {\rm{Re}}\left[\tilde{y}_{4,5}\right],
\,\quad {\rm{Im}}\left[\tilde{y}_{4,5}\right] \,\quad \},
\eea which stem from {
the terms
\bea \label{mfvpar30}
& &\widetilde{M}_1=e^{i\phi_1} M_1, \,\quad \widetilde{M}_2=e^{i \phi_2} M_2, \,\quad M_3, \,\quad
\tilde\mu=\mu e^{i\phi_\mu}, \,\quad M_A, \,\quad \tan \beta, \,\quad \nonumber\\ \nonumber
& &M^2_Q = {a}_1\  \One + x_1 X_{13} + y_1 X_1,
\quad \quad 
X_1 = {\rm{diag}}\{0,0, \delta_{3i} \delta_{3j} \},  
\quad \quad  X_{13} = V^*_{3i} V_{3j},\\ \nonumber
& &M^2_U = {a}_2 \ \One+ x_2\ X_1, \\\nonumber
& &M^2_D = {a}_3\ \One + y_3\ X_1, \\
& &M^2_L = {a}_6\ \One + y_6\ X_1, \\\nonumber
& &M^2_E = {a}_7\ \One + y_7\ X_1, \\\nonumber
& &A_E = \tilde{a}_8\ X_1, \\\nonumber
& &A_U = \tilde{a}_4\ X_5 + \tilde{y}_4\ X_1, \quad \quad X_5 = \delta_{3i} V_{3j}\\\nonumber
& &A_D = \tilde{a}_5 \ X_1 + \tilde{y}_5\ X_5,
\eea
  Here $i$ and $j$ run over as sparticles family indices, $V$ is
  the SM CKM matrix,  $\One$ the unit matrix 
   and $\delta$ the Kronecker delta function. 
The gaugino mass parameters $\widetilde{M}_1$, $\widetilde{M}_2$ were allowed in the
range -4 to 4 TeV for both real and imaginary parts. The gluino mass
term, $M_3$, is allowed in 100 GeV to 4 TeV. 
The parameters
 $~{a}_{1,2,3,6,7}$ were 
varied within the range $(100 \gev)^2$ to $(4 \tev)^2$ and $-(4
\tev)^2$ to $(4 \tev)^2$ for $x_{1,2}, y_{1,3,6,7}$. The trilinear
coupling terms 
Re$[\tilde{a}_{4,5,8}]$, Im$[\tilde{a}_{4,5,8}]$, Re$[\tilde{y}_{4,5}]$,
and ${\rm{Im}}(\tilde{y}_{4,5})$ 
 were varied within $-8 \tev$ to $8 \tev$. Here $m_A$ was
allowed in $100 \gev$ to $4 \tev$ while the Higgs doublets mixing
term, both real and imaginary parts ( 
${\rm{Re}}[\tilde\mu], \, {\rm{Im}}[\tilde\mu]$) in the range -4 to 4
TeV.  In comparison, the pMSSM 
 parameters are 
\bea \label{20par}
\underline{\theta} = \{ M_{1,2,3};\;\; m^{3rd \, gen}_{\tilde{f}_{Q,U,D,L,E}},\;\; 
m^{1st/2nd \, gen}_{\tilde{f}_{Q,U,D,L,E}}; \;\;A_{t,b,\tau,\mu=e},
\;\;m^2_{H_{u,d}}, \;\;\tan \beta\}, 
\eea
where $M_{1,2,3}$ are as for MSSM-30, $m_{\tilde f}$ the sfermion mass
parameters were allowed in the range 100 GeV to 4 TeV. The trilinear
couplings $A_{t,b,\tau,\mu=e} \in [-8, 8]$ TeV. The Higgs doublet
masses $m^2_{H_1}$, $m^2_{H_2}$ were allowed according to
$m^2 \in sign(m)\,[-4, 4]^2 \textrm{ TeV}^2$. Here $sign(\mu)$ is the sign
of the Higgs doublets mixing parameter (allowed to be randomly
$\pm 1$).  The SM parameters were fixed at their experimentally
determined central values for the  MSSM-30 but varied in a Gaussian manner
for the pMSSM.  

\subsection{The MSSM-30 global fit to data} 
The posterior distribution used for our analysis came from a Bayesian
fit of the MSSM-30 to data~\cite{AbdusSalam:2014uea}. 
In order to make this paper a self-contained exposition of the statistical details of our work, in what follows we describe the fitting procedure.

The Bayesian fit was performed within a context, $\cal{H}$, where the MSSM-30
neutralino lightest supersymmetric particle is assumed to be a least part of the cold dark matter (CDM)
relic. The thirty parameters, detailed in \eq{30parameters}, 
were varied according to a flat prior probability density, 
$p(\underline \theta | {\cal{H}})$.  The SM parameters fixed were: the mass of the Z-boson, $m_Z = 91.2 \gev$, 
the top quark mass, $m_t = 165.4 \gev$, the bottom quark mass, $m_b = 4.2
\gev$, the electromagnetic coupling, $\alpha_{em}^{-1} = 127.9$,
and the strong interaction coupling, $\alpha_s = 0.119$.

The data set, $\underline d$, used for fitting the MSSM-30 are summarised
in ~\Tabref{tab:obs}. It is composed of the experimental central
values, $\mu_i$, and errors, $\sigma_i$, for the Higgs boson mass, the
electroweak physics, B-physics, dipole moment of leptons and 
 the CDM relic density observables set
\bea
\underline O &\equiv &\{ m_h, \; m_W,\; \Gamma_Z,\; \sin^2\,
\theta^{lep}_{eff},\; R_l^0,\; R_{b,c}^0,\;
A_{FB}^{b,c},\;  A^l = A^e,\; A^{b,c}, \\ \nonumber
& &BR(B \rightarrow X_s \, \gamma),\; BR(B_s \rightarrow \mu^+ \,
\mu^-),\; \Delta M_{B_s},\;  R_{BR(B_u \rightarrow \tau \nu)},\\ \nonumber
& &\Omega_{CDM}h^2, \; Br(B_d \rightarrow \mu^+ \mu^-), \; \Delta
M_{B_d}, \; d_{e,\mu,\tau} \}.
\eea

\begin{table}[htbp!] 
\begin{tabular}{|ll||ll|}\hline
Observable & Constraint & Observable & Constraint  \\
\hline
$m_W$ [GeV]& $80.399 \pm  0.023$ \cite{verzo}&$A^l = A^e$& $0.1513 \pm
0.0021$ \cite{:2005ema} \\
$\Gamma_Z$ [GeV]& $2.4952 \pm 0.0023$ \cite{:2005ema}&$A^b$ & $0.923
\pm 0.020$ \cite{:2005ema}\\
$\sin^2\, \theta_{eff}^{lep}$  & $0.2324 \pm 0.0012$ \cite{:2005ema}&$A^c$ & $0.670 \pm 0.027$ \cite{:2005ema}\\
$R_l^0$ & $20.767 \pm 0.025$ \cite{:2005ema} &$\rm{BR}(B_s \rightarrow
\mu^+ \mu^-)$ & $3.2^{+1.5}_{-1.2} \times 10^{-9}$
\cite{Aaij:2012nna}\\
$R_b^0$ & $0.21629 \pm 0.00066$ \cite{:2005ema}&$\Delta M_{B_s}$ &
$17.77 \pm 0.12$  ps$^{-1}$ \cite{Abulencia:2006ze}\\
$R_c^0$ & $0.1721 \pm 0.0030$ \cite{:2005ema}&$R_{Br(B_u \rightarrow \tau \nu)}$&
$1.49 \pm 0.3091$ \cite{Aubert:2004kz}\\
$A_{\textrm{FB}}^b$ & $0.0992 \pm 0.0016$ \cite{:2005ema}& $\Delta
M_{B_d}$ & $0.507 \pm 0.005$ ps$^{-1}$\cite{Barberio:2008fa} \\
$A_{\textrm{FB}}^c$ & $0.0707 \pm 0.0035$ \cite{:2005ema}&$\Omega_{CDM} h^2$ & $0.11
\pm 0.02 $ \cite{0803.0547}\\
$m_h$ [GeV] & $125.6 \pm 3.0$ \cite{ATLAS:2013mma, CMS:yva} &  $\rm{BR}(B_d
\rightarrow \mu^+ \mu^-)$ & $<1.8 \times 10^{-8}$ \cite{Aaij:2011rja}\\
$d_\mu$ &$<2.8\times 10^{-19}$ \cite{McNabb:2004tj}& $\rm{BR}(B\rightarrow
X_s \gamma)$ & $(3.52 \pm 0.25) \times 10^{-4}$ \cite{Barberio:2007cr}\\
$d_\tau$ &$<1.1\times 10^{-17}$ \cite{Nakamura:2010zzi}
&$d_e$ &$<1.6\times 10^{-27}$ \cite{Regan:2002ta}\\
\hline
\end{tabular}
\caption{The experimental results used for the Bayesian fit of the
  MSSM-30 parameters. \label{tab:obs}}
\end{table}

Using the data set described above, an MSSM-30 likelihood distribution
$p(\underline d|\underline \theta, {\cal{H}})$ was constructed as
\be p(\underline d|\underline \theta, {\cal{H}}) = L(x) \,
\prod_i \, \frac{ \exp\left[- (O_i - \mu_i)^2/2
    \sigma_i^2\right]}{\sqrt{2\pi \sigma_i^2}},
\ee where the index $i$ runs over the list of observables $\underline
O$, the variable $x$ represents the predicted value of neutralino CDM
relic density at an MSSM-30 parameter space point and
\be \label{olik}
L(x) =
\begin{cases}
  1/(y + \sqrt{\pi s^2/2}) &  \textrm{ if } x < y \\
  \exp\left[-(x-y)^2/2s^2\right]/(y + \sqrt{\pi s^2/2}) &
  \textrm{ if } x \geq y
\end{cases}.
\ee
Here $y = 0.11$ is the CDM relic density central value and $s=0.02$
the corresponding inflated (to allow for theoretical uncertainties)
error. 

By passing parameters to {\sc SPHENO}~\cite{Porod:2003um,
  Porod:2011nf} 
   via the SLHA2~\cite{Allanach:2008qq} interface, the
corresponding MSSM-30 predictions for the branching ratios
BR$(B_s \rightarrow \mu^+\mu^-)$, BR$(B \rightarrow s \gamma)$,
$R_{\rm{BR}(B_u \rightarrow \tau \nu)}$, BR$(B_d \rightarrow \mu^+ \mu^-)$,
$\Delta M_{B_s}$,  $\Delta M_{B_d}$ and $d_{e,\mu,\tau}$ were
obtained. Similarly, using the SLHA1~\cite{Skands:2003cj} interface,
the neutralino CDM relic density 
was computed using {\sc
  micrOMEGAs}~\cite{Belanger:2008sj}, while
{\sc susyPOPE}~\cite{Heinemeyer:2006px,Heinemeyer:2007bw}
was used for computing precision observables that include the $W$-boson
mass $m_W$, the effective leptonic mixing angle variable
$\sin^2 \theta^{lep}_{eff}$, the total $Z$-boson decay width,
$\Gamma_Z$, and the other electroweak observables whose experimentally 
determined central values and associated errors are summarised in
\Tabref{tab:obs}.

Using {\sc  MultiNest}~\cite{Feroz:2007kg,Feroz:2008xx} which
implements the Nested Sampling algorithm~\cite{Skilling}, Bayes' theorem
then gives the MSSM-30 posterior probability distribution  
\be \label{bayes}
p(\underline \theta |\underline d, {\cal{H}}) \propto p(\underline d|\underline
\theta,{\cal{H}}) \, \times \, p(\underline \theta|{\cal{H}}).
\ee

\subsection{$\BRBsm$ prediction and measurement \label{BRBpredictionmeas}}

The tagged average branching fraction of the rare decay $\BRBsm$ is given by
\bea
\BRBsm&=&\frac{G_F^2\alpha^2}{64\pi^3} f^2_{B_s}m^3_{B_s}|V_{tb}V_{ts}^*|^2\tau_{B_s}
\sqrt{1-\frac{4m_\mu^2}{m_{B_s}^2}}\times\nonumber \\
& &\label{eq:thBR}\hspace*{-2cm}  \left[
  \left(1-\frac{4m_\mu^2}{m_{B_s}^2}\right)\frac{m^2_{B_s}}{m^2_b}|C_{S}-C_{S}'|^2 +
  \left| \frac{m_{B_s}}{m_b} ( C_P-C_P' ) + 2(C_{10}-C_{10}' )\frac{m_\mu}{m_{B_s}} \right|^2
  \right],
\eea
where the operators that we use above are related to those of  \cite{Chankowski:2000ng} by
\bea
\label{eq:WCoefbBRBs}
&& C_S= X (C^S_{LL}+C^S_{LR}),\quad \quad C_S'= X (C^S_{RR}+C^S_{RL}),\nonumber\\
&& C_P= X (-C^S_{LL}+C^S_{LR}),\quad C_P'= X (C^S_{RR}-C^S_{RL}),\nonumber\\
&& C_{10}= X (-C^V_{LL}+C^V_{LR}),\quad C_{10}'= X (C^V_{RR}-C^V_{RL}),
\eea
for $X=\pi/(\sqrt{2} G_F \alpha V_{ts}^* V_{tb})$,
and the 
Wilson operators and coefficients are defined through the Hamiltonian as 
\bea
 \mathcal{H}&=&-\sum_{X,Y} O^V_{XY} C^V_{XY}  + O^S_{XY} C^S_{XY},\nonumber\\
 O^V_{XY}&=&(d_J\gamma_\mu P_X d_I) (\ell_B\gamma^\mu P_Y\lambda_A),\nonumber\\
 O^S_{XY}&=&(d_J P_X d_I)(\ell_B P_Y\lambda_A), \quad X,Y=L,R,
\eea
where  $O^V$ are vector  and $O^S$ scalar operators respectively and $P_X$ are the chirality projectors.
 The contributions proportional to  $|C_{S}-C'_{S}|^2$ and  $|C_{P}-C_{P}'|^2$ are not any longer proportional to $m_\mu^2/m_{B_s}^2$ and hence lift the helicity suppression exhibited in the SM. 
 We use $\sflav$ \cite{Crivellin:2012jv} to obtain the contributions from the different particles and kinds of diagrams {\footnote{For this case, we use a modified version of the program, for which we have explicitly checked  those contributions with the help of references \cite{Babu:1999hn,Chankowski:2000ng,Logan:2000iv,Bobeth:2001sq,Bobeth:2002ch,Dedes:2008iw}.}}, Higgs and Z penguins and box diagrams.

It is well established that in the SM, $C_{10}$ gets its larger contribution from the $Z$ penguin with a top loop, about $75\%$ and its second largest contribution from the $W$ box, $24\%$ and we have
\bea
\BRBsm^{\rm{SM}}=(3.25\pm 0.17)\times 10^{-9}.
\eea
The experimental measured quantity (denoted here with an overline) is the untagged branching fraction which is related to 
\eq{eq:thBR}, the theoretical (tagged) expression,  as 
\bea
\label{eq:untaggedBR}
\BRBsm=\left[ \frac{1-y^2_s}{1+\ADF y_s}
  \right]
\Bsuntg,
\eea
where $y_s=\Delta \Gamma_s/2 \Gamma_s$, $\Delta \Gamma_s$ being the decay width difference between the $B_s$ mass eigenstates and $\Gamma_s=\tau^{-1}_{B_s}$ is the average $B_s$ decay width, using the LHCb measurement ($y_s=0.087\pm 0.014$ \cite{Raven:2012fb}), and that in the SM $\ADF$=1, we obtain{\footnote{
We note that this value is in agreement with \cite{Buras:2013uqa}, but a better treatment of NLO EW corrections to $C_{10}$ place $\Bsuntg^{\text{SM}}=(3.65\pm 0.23)\times 10^{-9}$ \cite{Bobeth:2013uxa}. }}
}
\bea
\label{eq:expBR}
\Bsuntg^{\text{SM}}=(3.56\pm 0.18)\times 10^{-9}, 
\eea
on the other hand, the experimental value measured by the LHCb
collaboration is \cite{Mulder:2017hug,Aaij:2017vad}, including Run 1 and Run 2 data,
\bea
\overline{\text{BR}}(B_s\rightarrow\mu^+\mu^-)=(3.0\pm 0.6^{+0.3}_{-0.2})\times 10^{-9}.
\eea
while the CMS value is \cite{Chatrchyan:2013bka},
\bea
\overline{\text{BR}}(B_s\rightarrow\mu^+\mu^-)=(2.8\pm 0.5^{+0.3}_{-0.2})\times 10^{-9}.
\eea
As we can see, both values in agreement with the SM.
In the MSSM,
$\ADF=(|P|^2\cos(2\varphi_P)-|S|^2\cos(2\varphi_S))/(|P|^2+|S|^2)$,
\cite{Buras:2013uqa}, where $\varphi_{S}=\arg(S)$, $\varphi_{P}=\arg(\! \, P)$, and $S$ and $P$ are related to our notation for the Wilson Coefficients as follows
\bea
S&=&\sqrt{1-4\frac{m_\mu^2}{m_{B_s}^2}} \frac{m_{B_s}^2}{2 m_\mu} \frac{1}{m_b+m_s}\frac{C_S-C_S'}{C_{10}^{SM}}{\frac{m_{B_s}}{m_b}}, \label{eq:CScont}\\ 
P&=&\frac{C_{10}}{C_{10}^{SM}}+ \frac{m_{B_s}^2}{2 m_\mu}\frac{1}{m_b+m_s} \frac{C_P-C_P'}{C_{10}^{SM}}{\frac{m_{B_s}}{m_b}}.
\label{eq:CPcont}
\eea

\begin{figure}[htp]
\centering
\includegraphics[width=8cm]{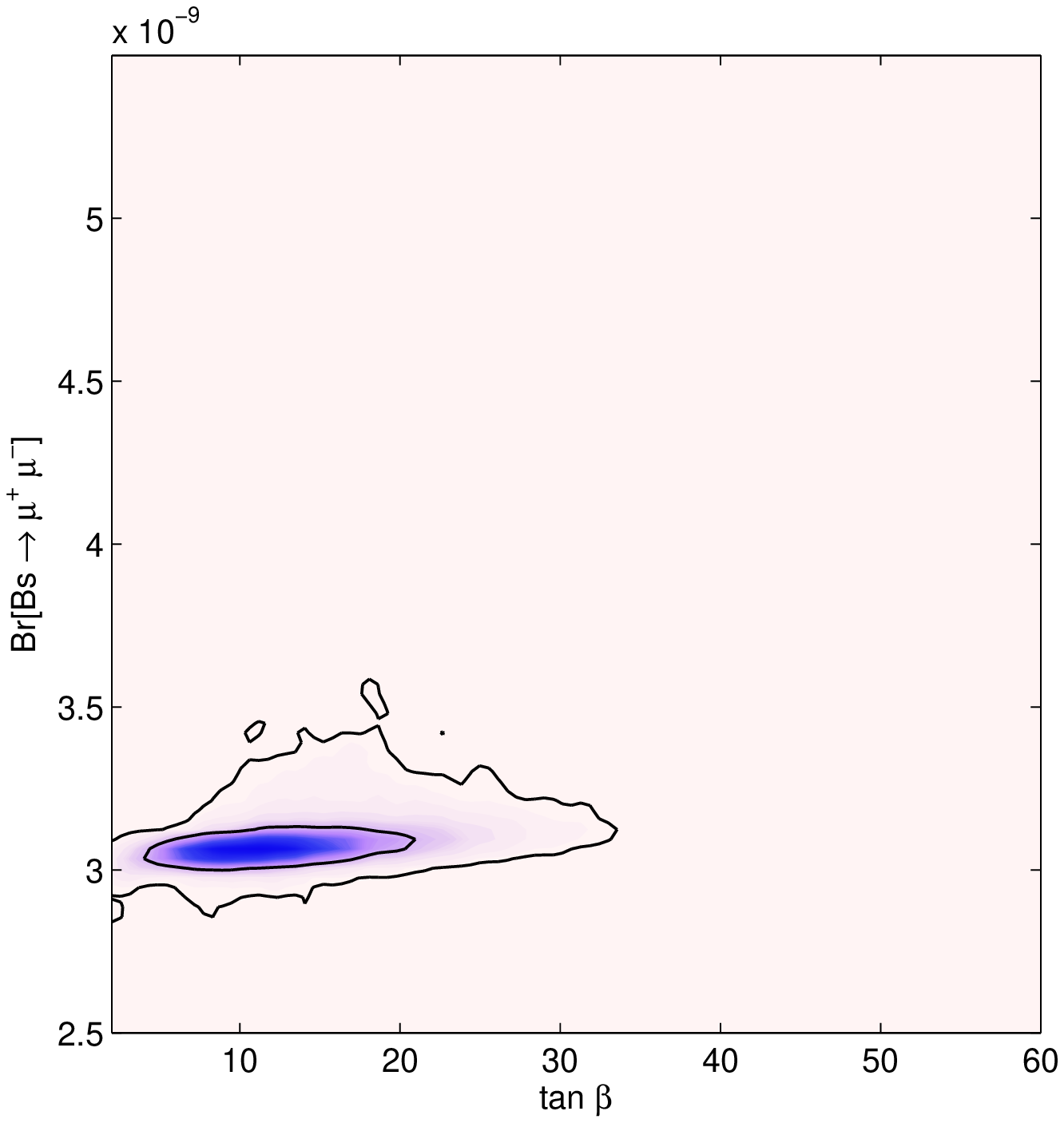}
\includegraphics[width=8cm]{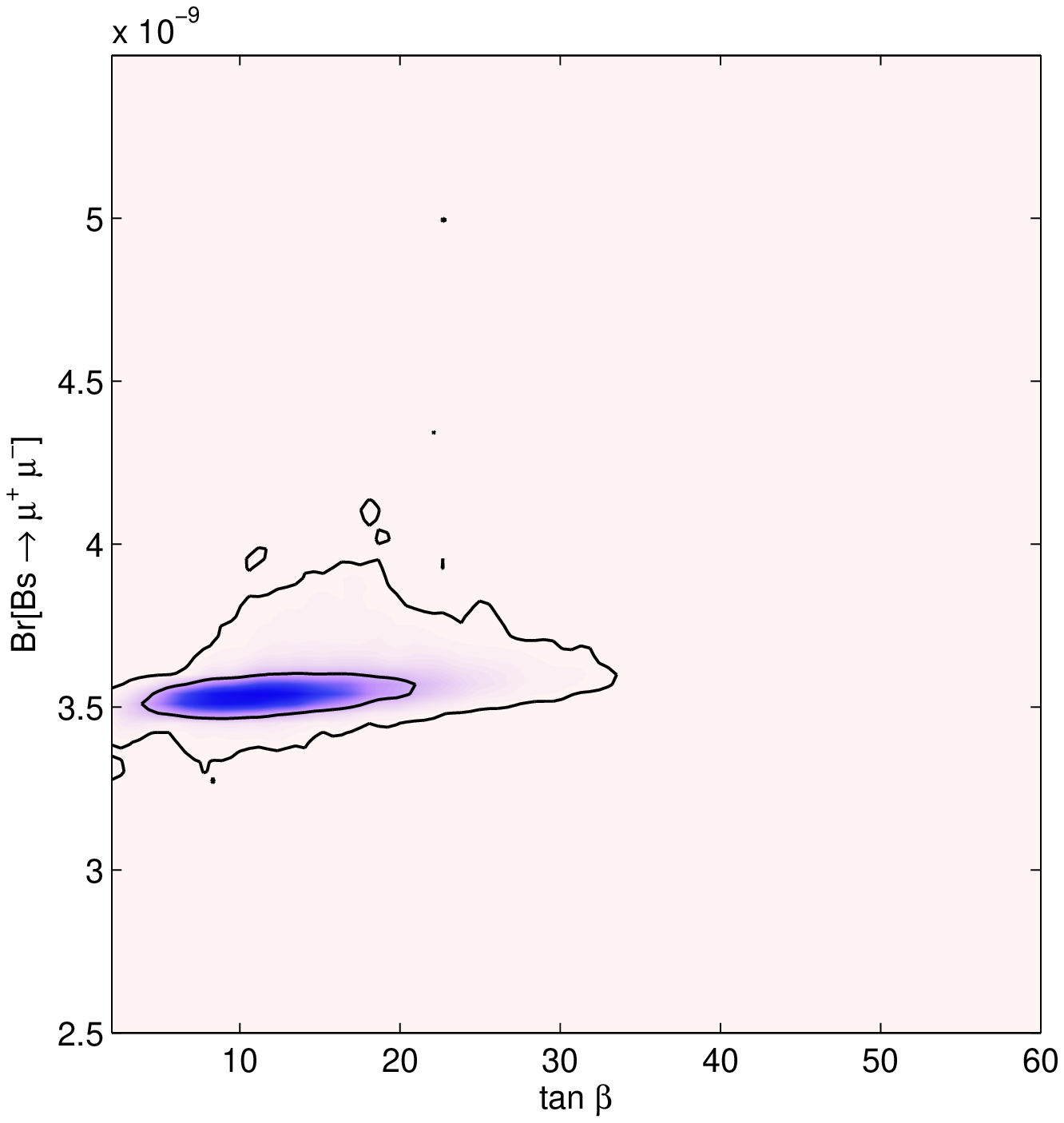}
\includegraphics[width=8cm]{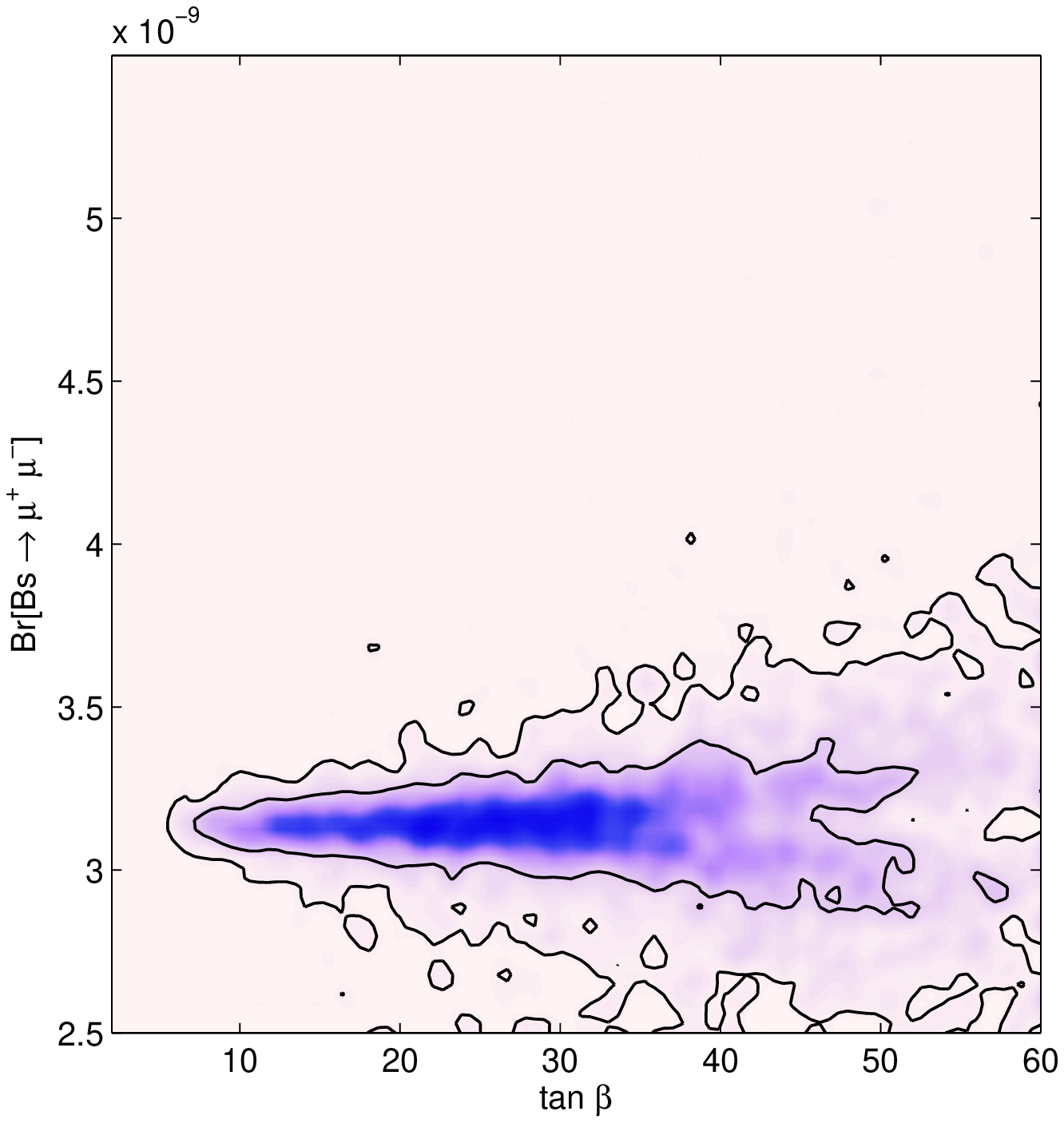}\\ 
\caption{$\tan\beta$ vs $\BRBsm$ for the MSSM-30 sample (top) and for
  the pMSSM (bottom). For the MSSM-30 sample we have plotted (left)
  the tagged contribution as produced by the official $2\_53$ version
  of $SUSY\_FLAVOR$ and the untagged values  (right) using a modified
  version of it. One of the main results of our work is to note that
  whereas values for 
  $\tan\beta<10$ are excluded in the pMSSM, for the MSSM-30 such low
  values of $\tan\beta$ are possible.  
\label{fig:tbvsBR}}
\end{figure}

 In the upper panels of \Figref{fig:tbvsBR}  we compare the planes $\tan\beta$ vs  the tagged value of  $\BRBsm$   (top-left),  as produced by the official $2\_53$ version of $SUSY\_FLAVOR$, and  $\tan\beta$ vs  the  untagged value of  $\BRBsm$  (top-right) using a modified version of it.  This comparison shows the importance of appropriately comparing the measurement of the  $\BRBsm$  with the theoretical value. Although the contributions from the pre-factors in \eq{eq:untaggedBR} do not differ greatly from point to point (due to the smallness of the supersymmetric contributions), they have a significant impact in pushing up the values of $\BRBsm$. In the lower part of the figure we present just the tagged distribution for the  linear prior of the pMSSM, which is in agreement with that of \cite{Arbey:2012ax}. 
 In the plane $\tan\beta$ vs $\BRBsm$ is clear that for the MSSM-30,
 contrary to the pMSSM, values of  $\tan\beta<10$ are not
 excluded. This shows that allowing a richer structure in the
 soft-squared terms, opens up regions of parameter space in comparison
 to the pMSSM. One of the main results of this work is that
 we found that the MSSM-30 predictions for $\BRBsm$ with $\tan\beta\in (10,20)$ are within the experimental limit, while for the pMSSM in that range
 are not. For the pMSSM instead the preferred values for $\tan\beta$ are above 25.
 
From the second plot of \Figref{fig:tbvsBR}, we can see that supersymmetric contributions add up little to the SM contribution, except for values between $\tan\beta \sim(10,20)$, where there could be both enhancing or suppressing effects. 
In \Secref{sec:Bsmupmum}, we present the numerical anatomy of the
$\BRBsm$ based on the MSSM-30 frame for which there are sources for CP violation beyond the CKM. 
We shall comment 
    on the interplay of the contributions coming
    from the neutral Higgs, $H^0$ and $Z$ penguin diagrams after
    introducing the theory of supersymmetric contributions to $C_S$, $C_P$
    and $C_{10}$.  
    The box diagram
    contribution to any of the Wilson Coefficients entering into $\BRBsm$ is
    small in comparison to the SM \cite{Chankowski:2000ng}. Although this is strictly true in the case where the CKM matrix is the only source of CP and flavor violation, in our case the contributions from the  extra sources of CP violation are generally negligible. We comment very briefly about the box contribution in Sec. \ref{Sec:Diag-PP}.

\begin{figure}[htp]
  \centering
  \includegraphics[width=7cm]{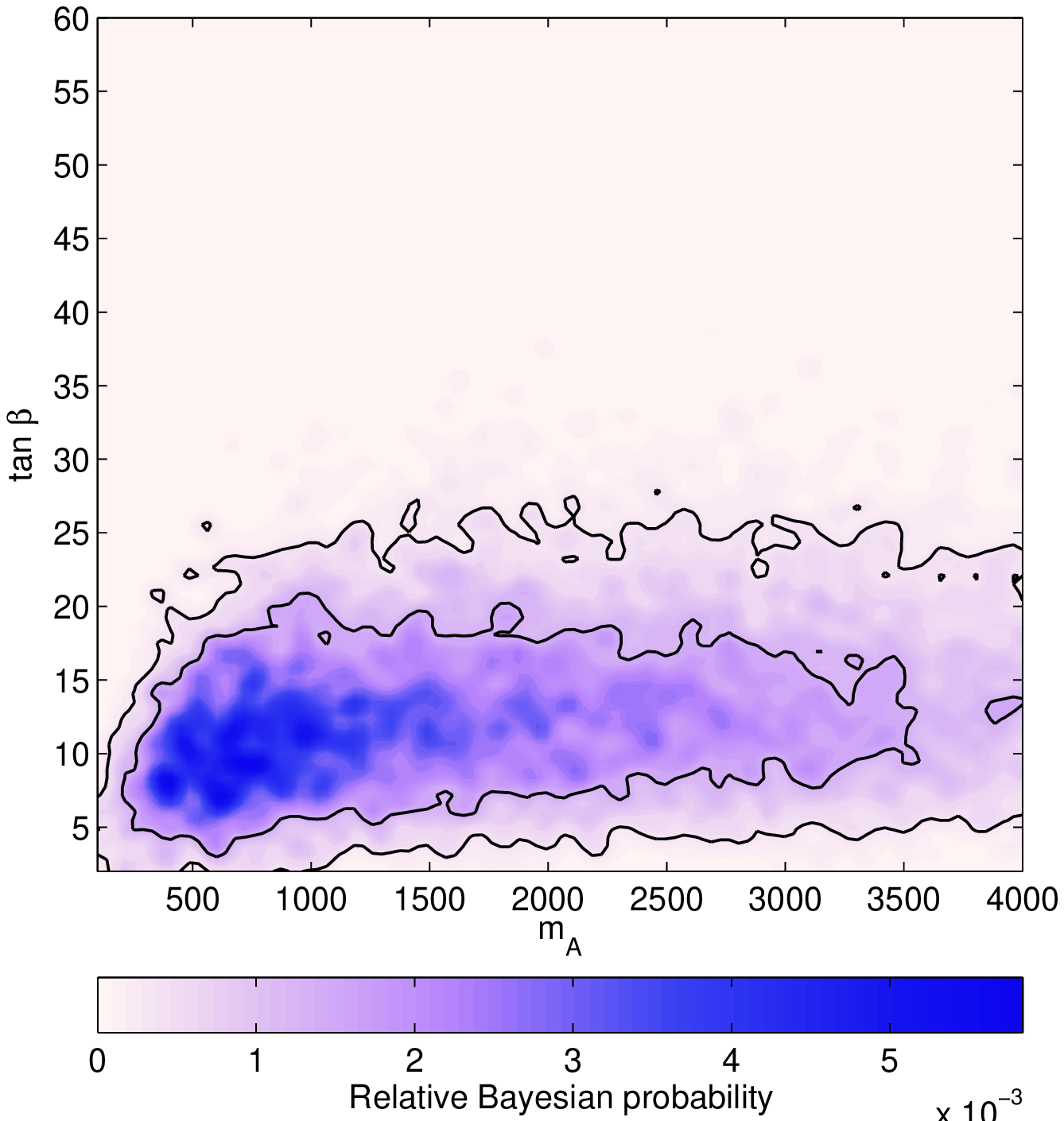}
  \includegraphics[width=7cm]{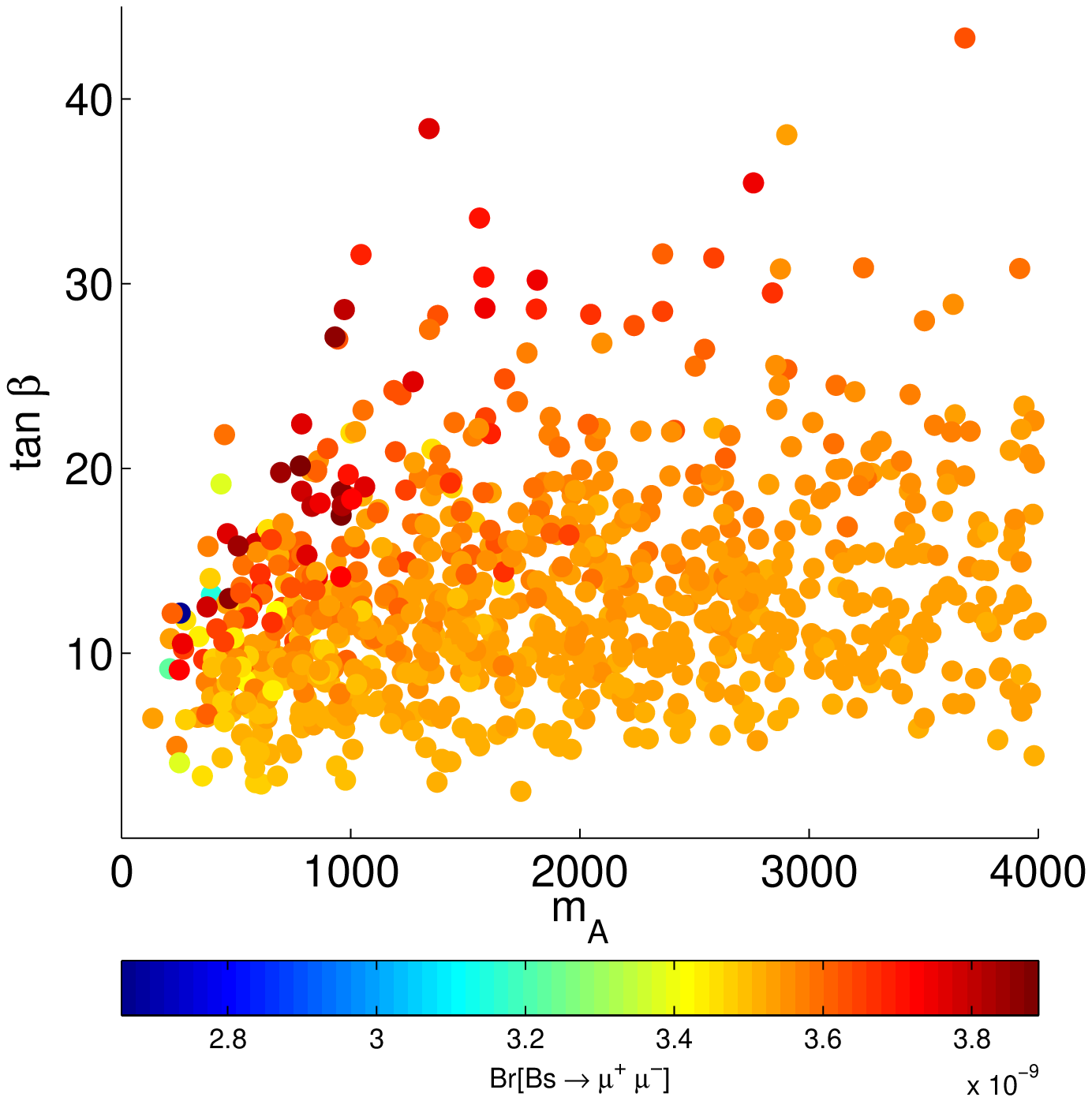} 
  \caption{Left: The marginalised 2D MSSM-30 $(m_A, \tan \beta)$
    posterior distribution. The outer and inner contours enclose the 
    $95\%$ and $68\%$ Bayesian probability regions respectively,  $m_A$
    is in a GeV scale. Right: The scatter plot on the same plane
    shows the  correlation to $\BRBsm$ without taking parameter points
    associated Bayesain posterior probalities. 
  }
  \label{fig:tbvsBR0}
\end{figure}

For this work, the relevant posterior probability distribution, of the
form of \eq{bayes}, which is a marginalised over the 2D $(m_A,
\tan \beta)$ plane  is shown in
\Figref{fig:tbvsBR0}. A 3D scatter plot showing the
variations of $\BRBsm$ on the same place is also shown. {One of 
  the aims of this article is to analyse the different contributions 
  to the $\BRBsm$ within the MSSM-30 posterior and to assess the
  impact of the recent $\BRBsm$ measurement on the MSSM-30 parameters 
  posterior. We shall address the latter case in what follows and the
  former in \Secref{sec:Bsmupmum}.} 
The posterior distribution in \Figref{fig:tbvsBR0} (left) shows that
high $\tan \beta$ values are disfavoured. The result of the MSSM-30
global fit shows that $\tan \beta$ lies in the range $4.5$ to $26.9$
at $95\%$ Bayesian probability interval. This feature is new compared to the
pMSSM fits in \cite{AbdusSalam:2008uv, AbdusSalam:2009qd}. The reason
why the two distributions have different shape 
is 
mainly due to the inclusion of new CP-violating parameters in the
MSSM-30 and the leptonic electric dipole moment constraints which tend
to be proportional to $\tan \beta$~\cite{Pospelov:2005pr}. 
In order to limit 
the over production of the dipole moments, 
relatively
lower, in comparison to the pMSSM, $\tan \beta$ values
are needed. In \Figref{fig:tbvsBR0} (right), the
weight-free scatter plot shows the correlations of $\BRBsm$ along the
$m_A$ or $\tan \beta$ directions within the global fit posterior. It
can be seen that independently of $m_A$ above some few 100s of GeV and
for $\tan \beta \sim 5$, the value of $\BRBsm \sim 3.5 \times 10^{-9}$  is
constant. This 
indicates a possible tension between the MSSM-30
global fit posterior described here with the $\BRBsm$
measurement~\cite{Mulder:2017hug} given that, for instance, assuming a
future $\BRBsm$ precision of $15\%$ 
 relative to the central value kills most of the posterior points and
 the surviving ones have sub-TeV $m_A$. This result is only
   indicative. A robust inference concerning the impact of such a
   plausible future $\BRBsm$ precision will require new fits of the
   MSSM-30 to data.
 This is because the result and any other feature within the
   posterior sample is obviously due to the resultant effect of
   the various observables 
in \Tabref{tab:obs} used  for 
 constraining the MSSM-30 parameters. The
main message here is that current $\BRBsm$
measurement~\cite{Mulder:2017hug} and possible future precisions
will most likely reduce the allowed MSSM-30 region.

\section{MSSM-30 contributions to $\Bsm$}
\label{sec:Bsmupmum}
For the MSSM-30 there are new sources of flavor and CP violation beyond  the CKM 
and therefore the  
contribution from different particles becomes relevant. As it will be
shown later, the neutralino and gluino contributions can compete with
those from charginos. {{Although} all of these contributions are suppressed in the
MSSM-30 posterior sample with $\tan \beta$ typically less than
30. Therefore, for making contrast to the various BSM contributions
within the pMSSM and MSSM-30, different regimes for $\tan\beta$ are
considered.

\subsection{Diagram-by-diagram and particle-by-particle contributions \label{Sec:Diag-PP}}
From here on, we refer to the Box, Higgs penguin and Z penguin
diagrams as kind of diagrams for which some examples are shown in \Figref{fig:feynmandiagms}.  As mentioned {in \Secref{BRBpredictionmeas}},  $C_{10}$  in the SM gets its larger contribution from the $Z$ penguin with a top loop, about $75\%$ and its second largest contribution from the $W$ box, $24\%$. Higgs penguin contributions in the SM are highly suppressed. On the other hand, the largest contribution in the pMSSM comes from the  second diagram of \Figref{fig:feynmandiagms} since  the degeneracy of scalar masses $\tilde{Q}$ is broken by radiative effects  induced by Yukawa couplings. This produces and effective flavor off-diagonal piece which does not go away when rotating to the mass eigenstates basis \cite{Babu:1999hn}. In the MSSM-30 off-diagonal elements are present and compete with the contribution coming from the afore mentioned radiative effects. In the case of MSSM-30, contributions from the $Z$ penguin diagrams are in general suppressed, except for low value of $\tan\beta$ ($\lesssim 10$).

\begin{figure}[htp]
  \centering
  \includegraphics[width=12cm]{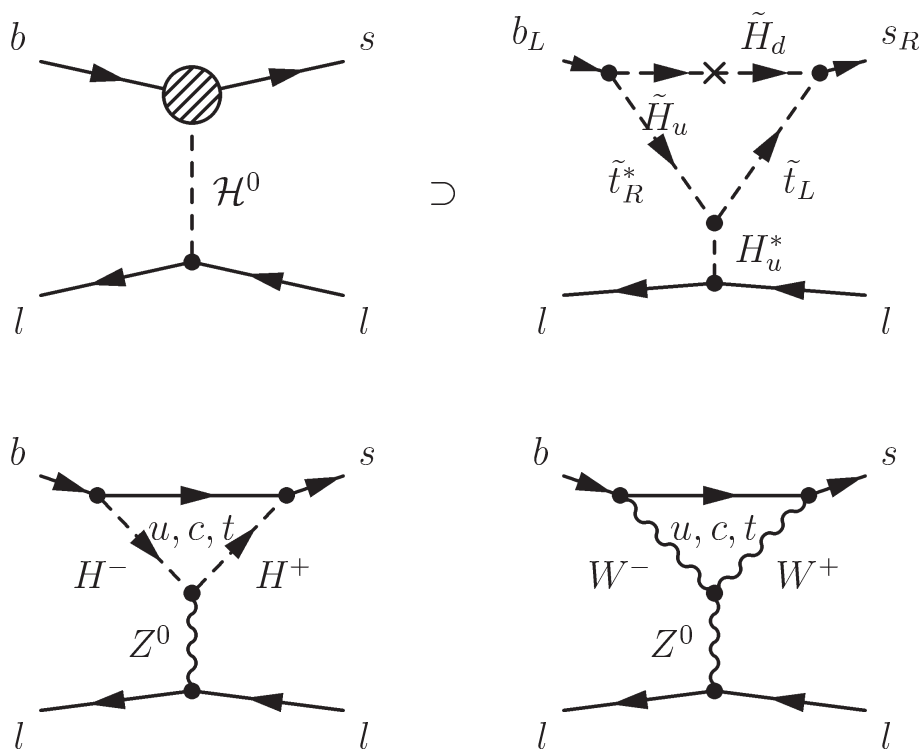}
  \caption{Higgs penguin and Z
    penguin diagrams  contributing to $\BRBsm$.  The first diagram represents the general Higgs penguin diagram. The most important contribution in the pMSSM comes from the second diagram of the first line, depicted in flavor basis. Even in the pMSSM this is the most important contribution since the degeneracy of scalar masses $\tilde{Q}$ is broken by radiative effects  induced by Yukawa couplings and hence induces and effective flavor off-diagonal piece which does not go away when rotating to mass eigenstates basis \cite{Babu:1999hn}.
    The last Z penguin gives the leading SM contribution.  \label{fig:feynmandiagms}}
\end{figure}

 In  \Figref{fig:Diagdecomposition} and \Figref{fig:ParticlesContr}
 we compare the pMSSM and the MSSM-30 in terms of their
 contributions to $\BRBsm$ coming from different diagrams and
 particles respectively. 
 Here by ``particles", we refer to ``gluino", ``chargino", ``neutralino" and ``W + charged Higgs boson", understanding that these particles can only come in their respective loops  together with squarks type down,  squarks type up,  squarks type down and quarks type up respectively. 
   The contributions for the pMSSM (solid
 black lines) are compared to the MSSM-30 case (red dashed lines). We
 can see that the  
distributions for the Z penguin and Box diagrams become narrower in
the MSSM-30, in comparison to those of the pMSSM, but overall these
contributions  shift  $\BRBsm$ to higher values than in the pMSSM
case. For the MSSM-30, the contributions to the Higgs-Penguin diagrams
become a bit suppressed, because of the preferred bigger masses for
$m_A$ and $m_{\tilde t}$ and lower values for $\tan\beta$. 
\begin{figure}[htp]
  \vspace{0.5cm}
  \centering
  \includegraphics[width=6cm]{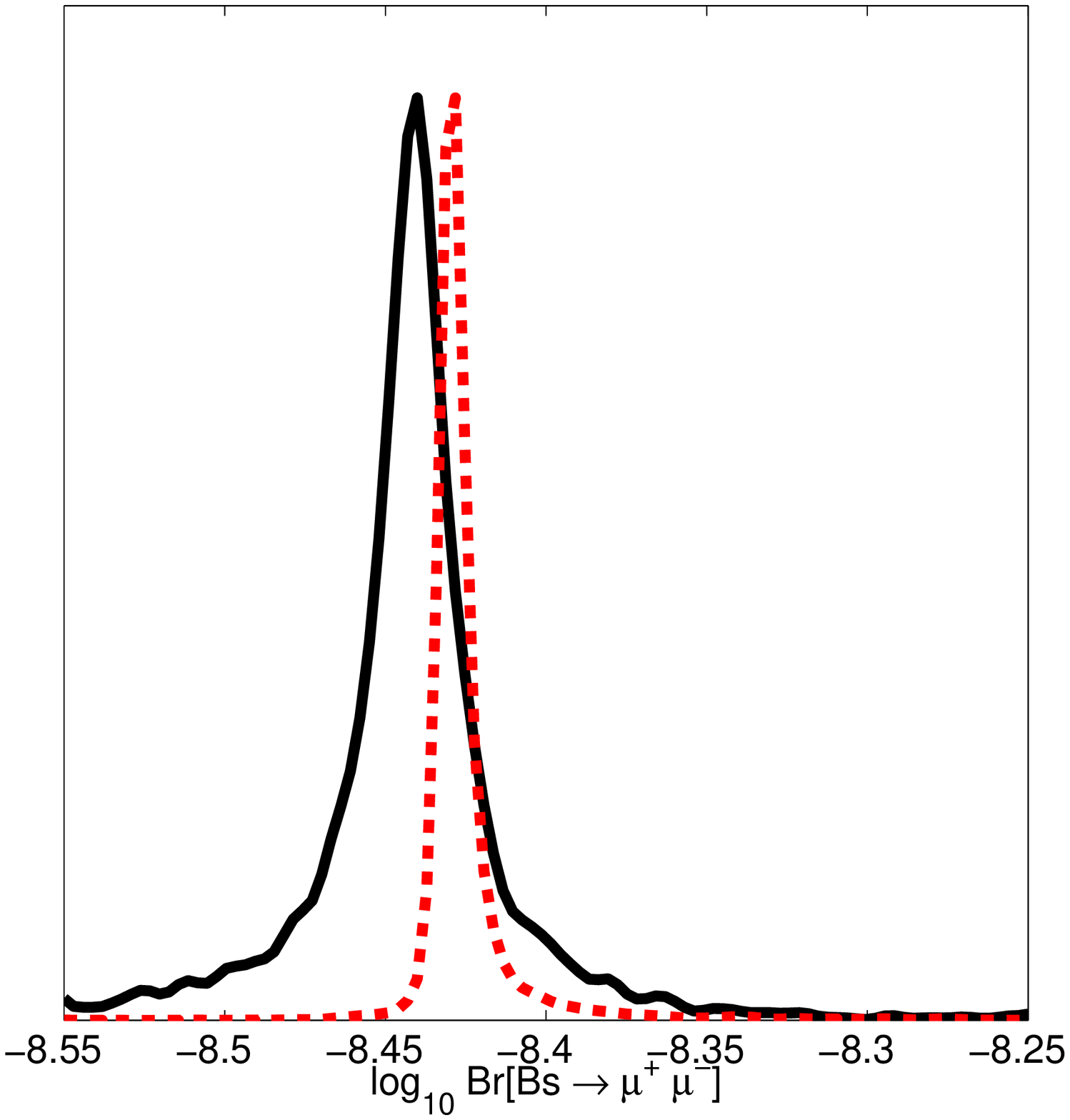}
  \includegraphics[width=6cm]{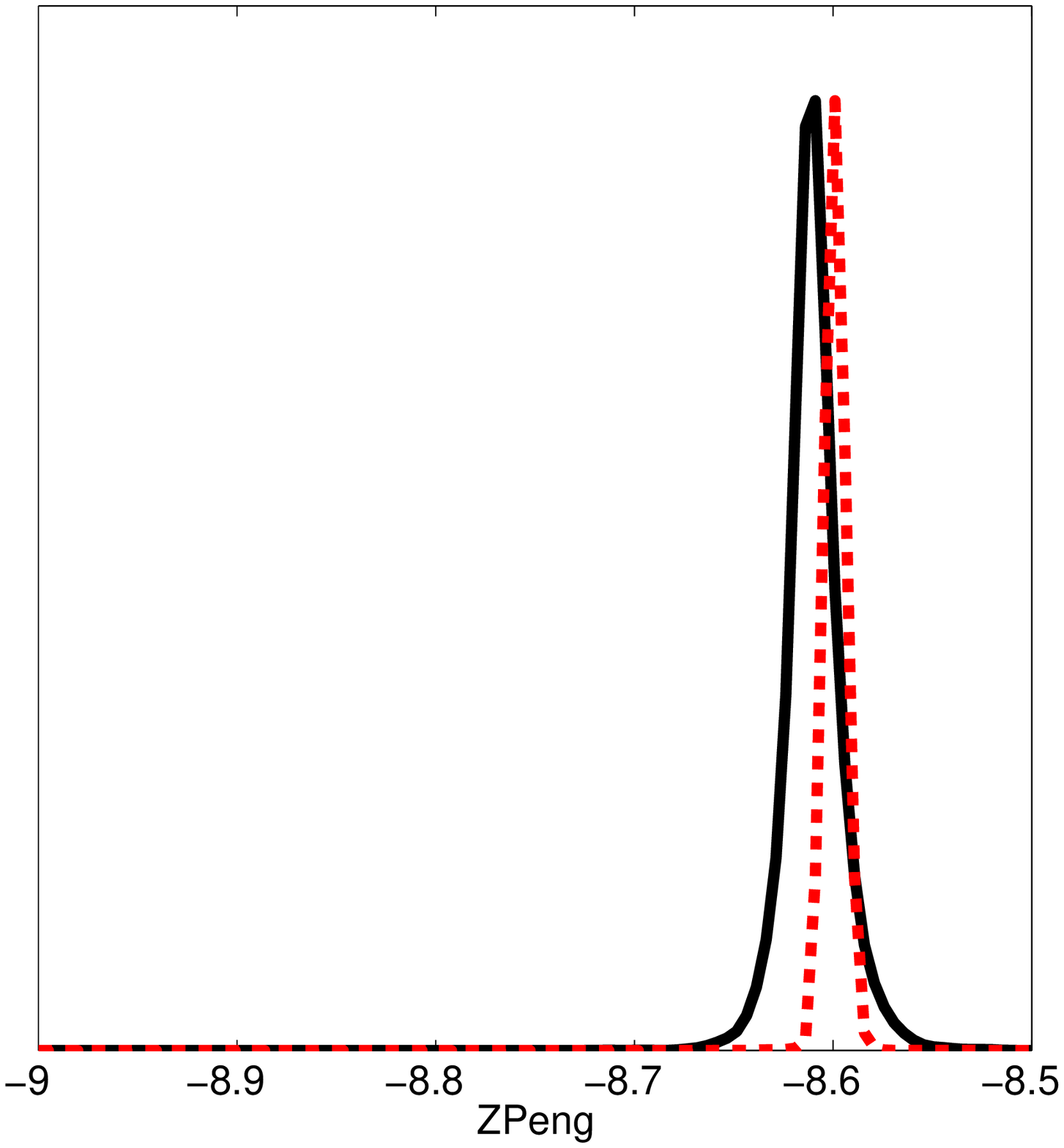}\\
  \includegraphics[width=6cm]{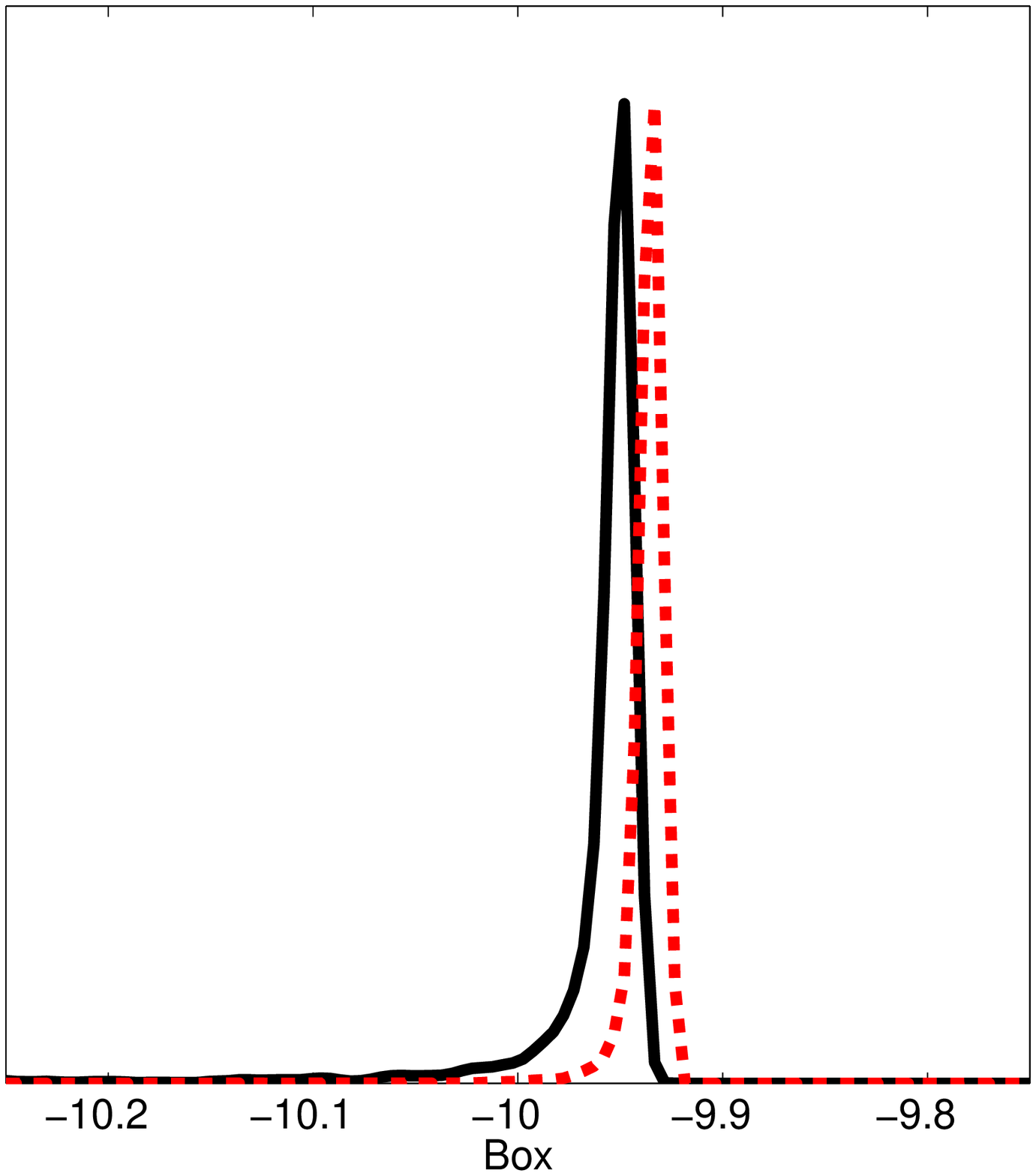}
  \includegraphics[width=6cm]{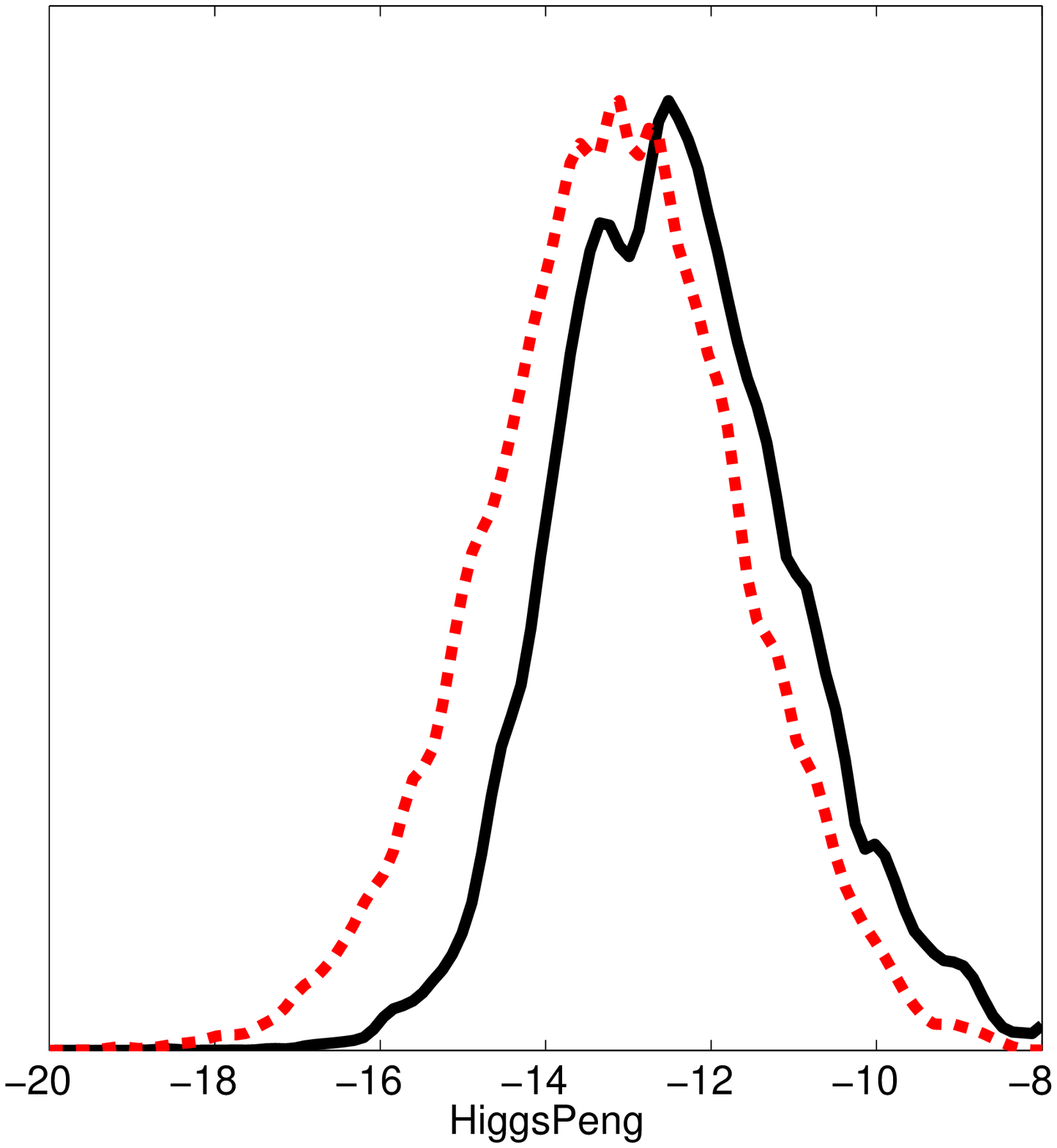}
  \caption{Contribution by diagrams to $\BRBsm$. The solid black lines
    correspond to the pMSSM and the dashed red lines to the MSSM-30.
    The distributions for the Z penguin and Box diagrams become
    narrower in the MSSM-30 in comparison to those of the pMSSM, but
    overall these contributions shift  $\BRBsm$ to higher values than
    in the pMSSM case.  For the MSSM-30, the contributions to the
    Higgs-Penguin diagrams  become suppressed, due to the preferred
    bigger masses for $m_A$ and $m_{\tilde t}$ and lower values for
    $\tan\beta$. For all plots, the vertical axis represent the
      relative probability density associated to a point in the MSSM scan.
      The horizontal axis is the logarithm of the corresponding
      contribution to $\BRBsm$. \label{fig:Diagdecomposition}}  
\end{figure}

\begin{figure}[htp]
  \centering
  \includegraphics[width=6cm]{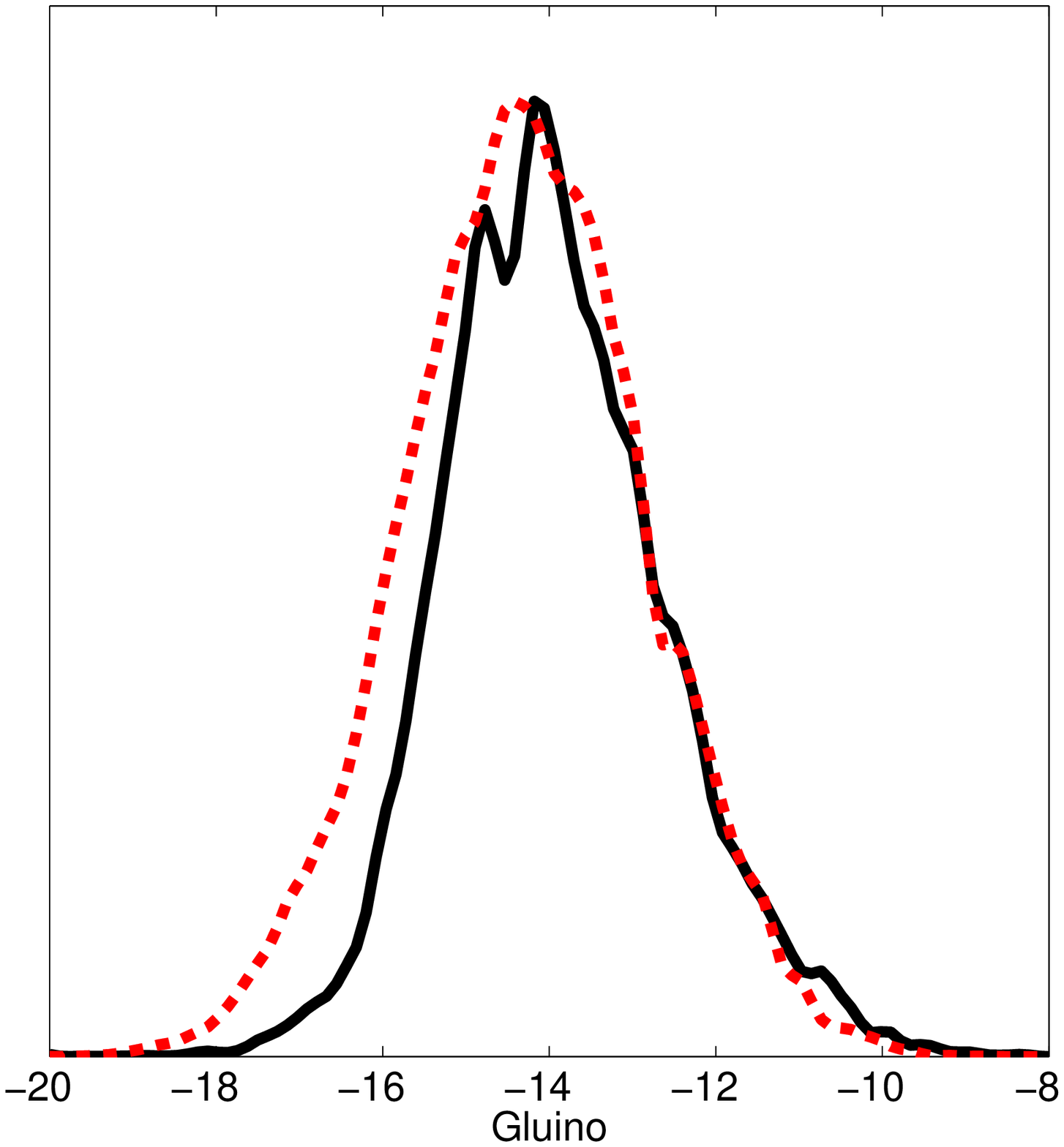}
  \includegraphics[width=6cm]{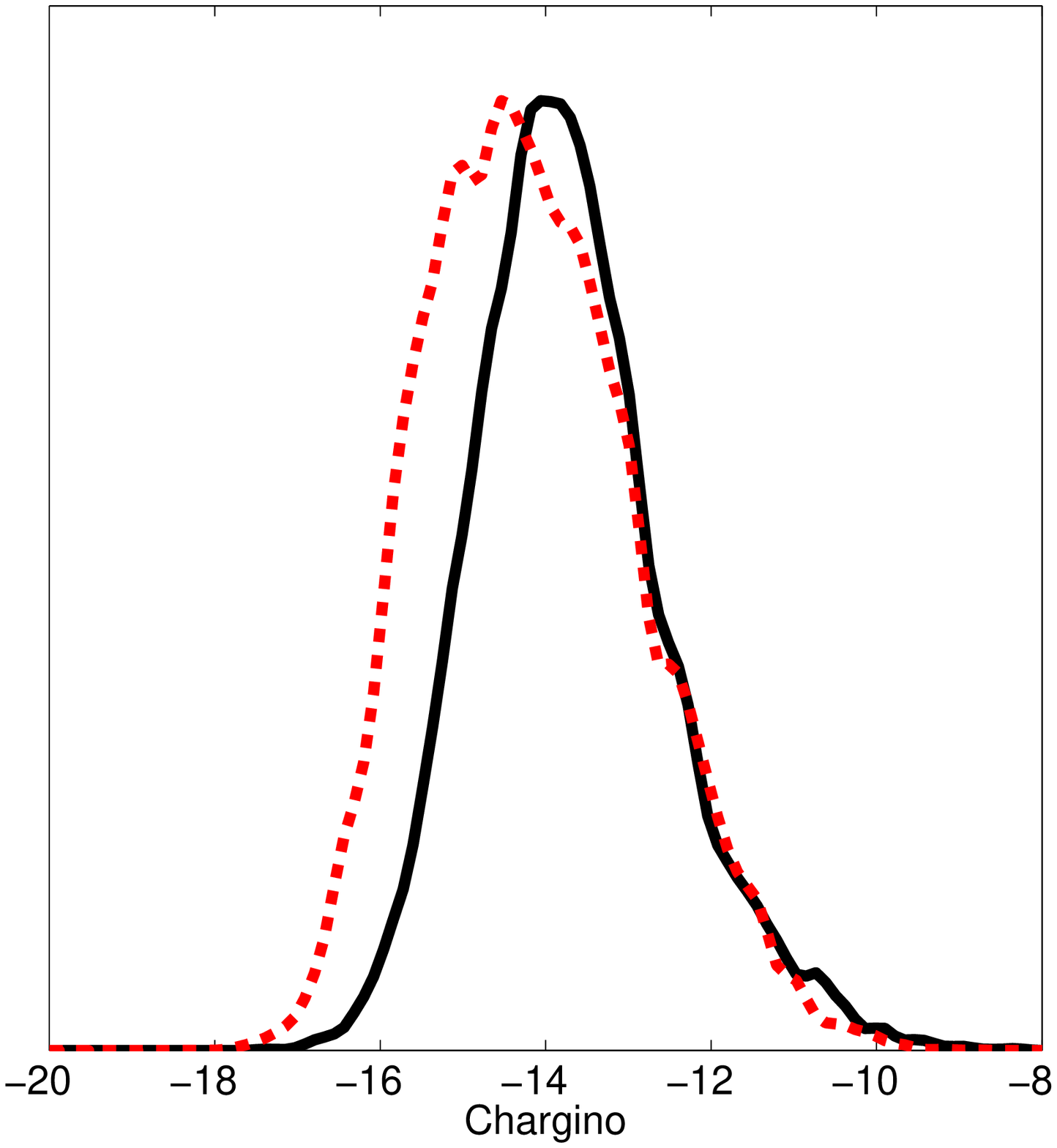}\\
  \includegraphics[width=6cm]{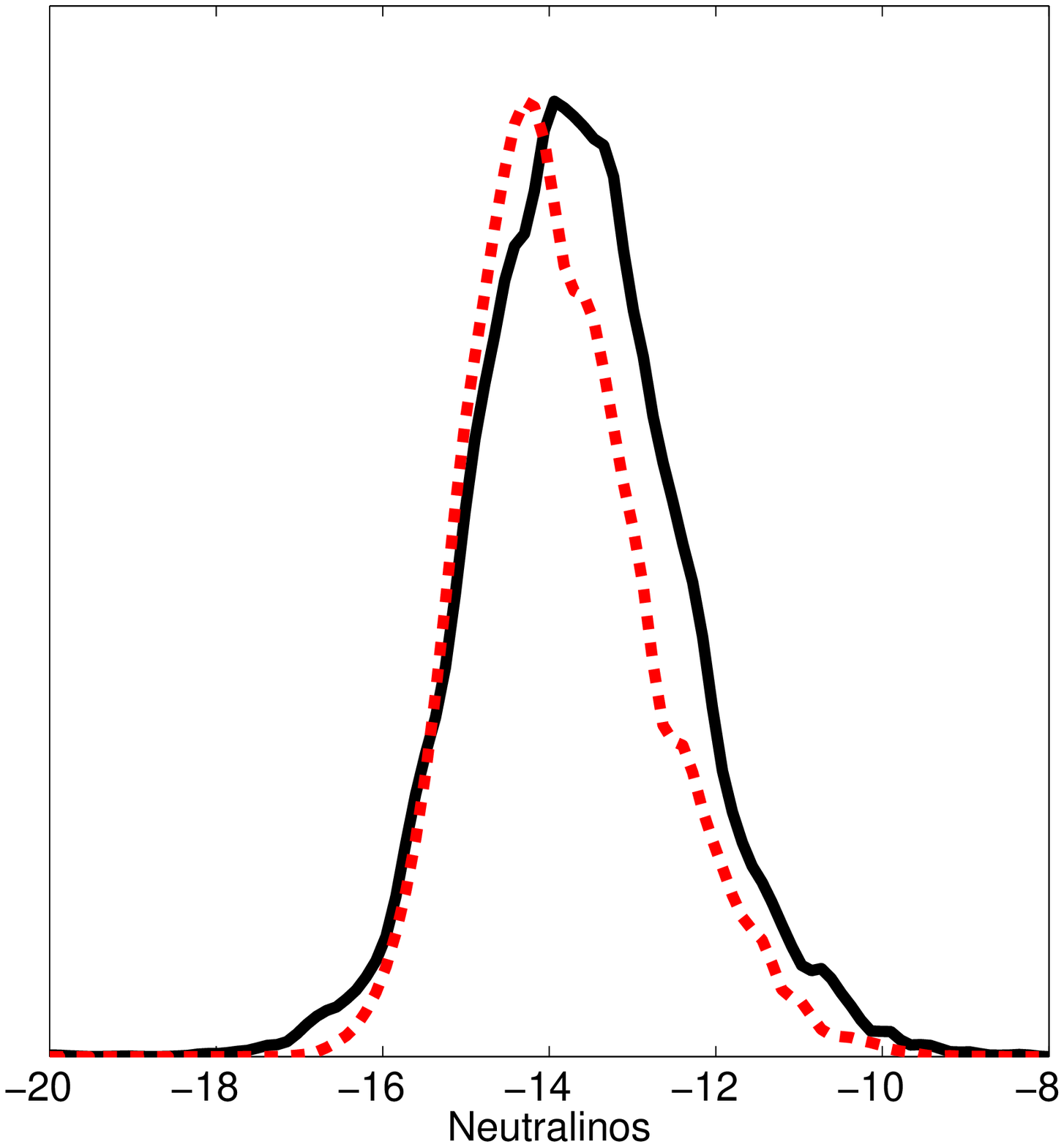}
  \includegraphics[width=6cm]{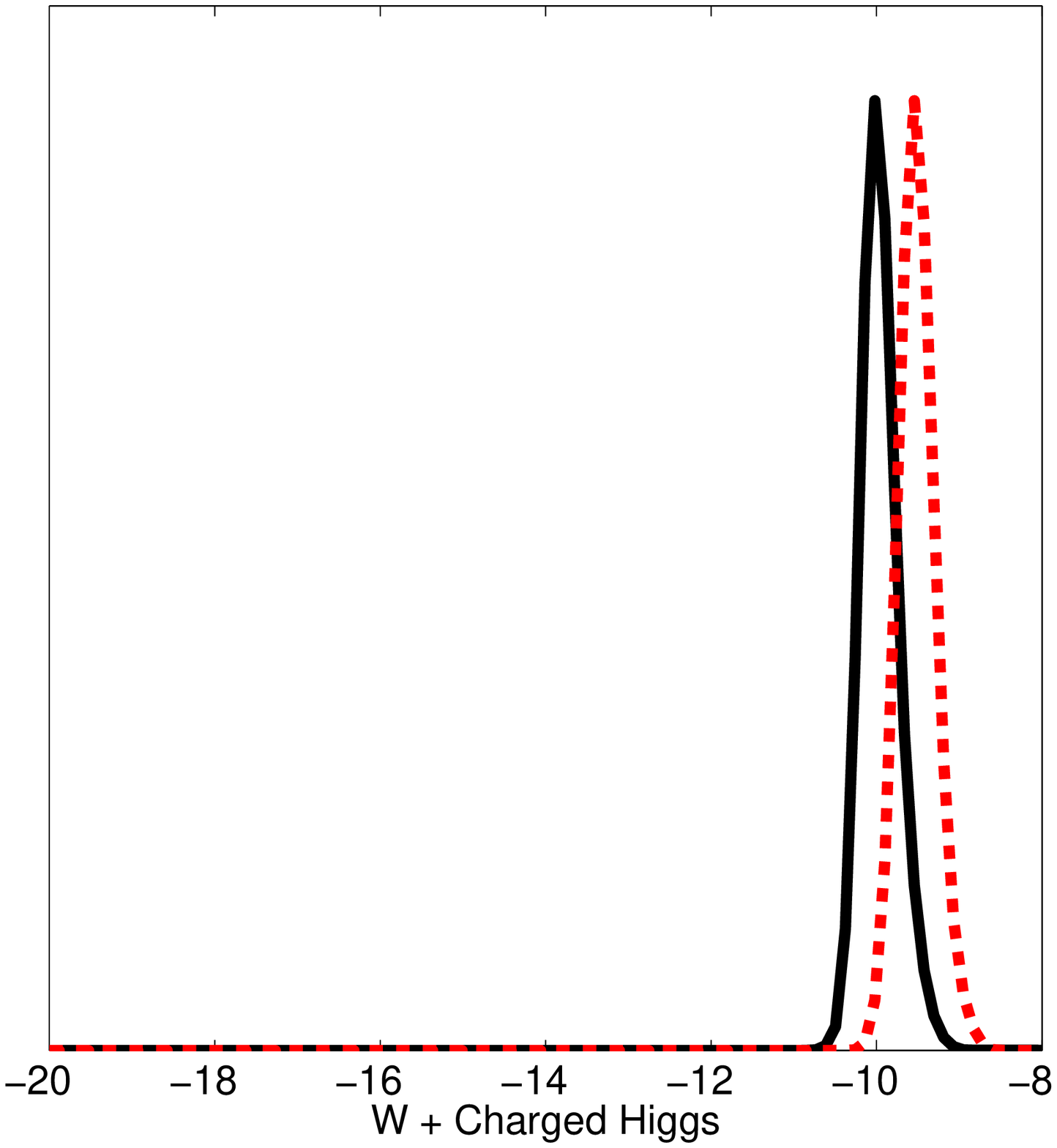}
  \caption{The particle-by-particle contributions to  $\BRBsm$. The solid black lines 
    correspond to the pMSSM and the dashed red lines to the
    MSSM-30. In $\sflav$ the charged-Higgs contribution cannot be
    separated from the SM  W-boson contribution. The axes are as in Fig. 3.  \label{fig:ParticlesContr}}
\end{figure}

\subsection{Contributions to $C_P$ and $C_S$ \label{sbsc:contsCPCS}}

\subsubsection*{$H^0$-penguin \label{sec:H0penguin}}

As mentioned above, the most important contribution in the pMSSM comes from the Higgs penguin diagram depicted in the second diagram of \Figref{fig:feynmandiagms}. This  happens because the degeneracy of scalar masses $\tilde{Q}$ is broken by radiative effects  induced by Yukawa couplings and hence this induces and effective flavor off-diagonal piece which does not go away when rotating to mass eigenstate basis \cite{Babu:1999hn}.  The large $\tan\beta$ region of this contribution can play a very significant role. This can be understood by writing the simplified contribution at LO as \cite{Bobeth:2001sq}
\bea
\label{eq:FSandFP}
& &
C^{H^0}_S(\Cpm)\approx -C^{H^0}_P(\Cpm)\nonumber\\
& &= \mu A_t 
\frac{ \tan^3\beta}{(1+\epsilon_b \tan\beta)^2}
\frac{m_t^2}{m^2_{\tilde t}} \frac{m_b m_\mu}{4 M^2_W m^2_A\sin^2\theta_W}
x\left[\frac{(1-x)+\log(x)}{(1-x)^2} \right],\nonumber\\
& & 
x= m^2_{\tilde t}/m_{\Cpm_1}^2, \quad m_{\Cpm_1}\approx \mu,
\label{eq:pMSSMCsCp}
\eea
which is quite sensitive to $m_A$ and  $m_{\tilde t}$ and therefore drops noticeably with the increase of their values.  
From this expression, we can also understand that the lower the value of $\tan\beta$, the lower the contribution to $\BRBsm$ from this diagram. Since for our fits, the preferred values of $m_{\tilde t}$ are $O(1)$ TeV for the pMSSM,  the  suppression of this contribution becomes  considerable. 

In the MSSM-30, the off-diagonal parameters in the soft-squared masses and trilinear terms, \eq{30parameters}, add up to the contributions given by the broken degeneracy of the of scalar masses $\tilde{Q}$. In this case, the contributions to $C^{H^0}_{S,P}(\Cpm)$ cannot be written in the form of \eq{eq:pMSSMCsCp}, because non-zero off-diagonal terms are present even before the breaking of the degeneracy of the diagonal soft-squared masses. However, we find that the differences between the contributions of  $C^{H^0}_{S,P}(\Cpm)$ in the pMSSM and in the MSSM-30 is only at the percent level.  {In this respect, in the MSSM-30, there could be 
cancellations among these two contributions for equally heavy/high magnitude
parameters (e.g. $\mu$ and $A_t$).}
These cancellations could make the $\BRBsm$ independent of $m_A$ as shown in \Figref{fig:tbvsBR} (top-right), and  make the contribution to the Wilson Coefficients not so different from the pMSSM.

\subsubsection*{$Z$-penguin and Box diagrams}
 Supersymmetric particles propagating in the loop cannot generate a contributions to $C_{S,P}, C_{S,P}'$ due to the vector coupling to $Z^0$. Diagrams with charginos propagating in the box also give rise to non zero values of  $C^\chi_S$ and $C^\chi_P$ but leave $C^{'\chi}_S=C^{'\chi}_P=0$. In the case where the masses of squark and sneutrino in the box are degenerate, one can have $C^\chi_S=-C^\chi_P$ . Diagrams with one charged Higgs boson propagating in the box can give non-zero contributions only through the left handed parts so that   $C^{H^+}_S=-C^{H^+}_P$ and  $C^{' H^+}_S=-C^{' H^+}_P$ \cite{Becirevic:2012fy}.

\subsection{Contributions to $C_{10}$ \label{sec:susycont_toC10}}

\subsubsection*{$Z$-penguin} 

 For  heavy
  charged Higgs bosons and low values of $\tan\beta \leq 20$, Higgs
 penguin and box diagrams are small. Hence, the contribution from the
 $Z$ (third diagram in \Figref{fig:feynmandiagms}) penguin becomes the dominant one among all of the contributions to $C_{10}$.  
 This effect becomes accentuated for very small values of
 $\tan\beta$. 

 Our fits favor intermediate values of $\tan\beta$, so in principle there is a transition between the regimes of $H^0$ and $Z$ penguin dominance. However, this also depends on the values of the charged-Higgs mass and most importantly on the value of $m_{H}$, being large, there is not a surprise that both contributions are mostly suppressed. At LO, the corresponding contributions to $C_{10}$ and $C_{10}'$  are \cite{Bobeth:2001sq}
\bea
C_{10}^{Z}(H^{\pm})&=&  \frac{1}{8 \sin^2_W}\frac{m_t^2}{m_W^2}\frac{1}{\tan^2\beta}\ f_H(y_t), \nonumber\\
C_{10}^{' \ Z}(H^{\pm})&=&  - \frac{1}{8 \sin^2_W}\frac{m_s m_b}{m_W^2}\tan^2\beta \ f_H(y_t),\nonumber\\\
& &   f_H(y_t)=\frac{y_t}{1-y_t}\left( 1+ \frac{1}{1-y_t}\log y_t \right),\ y_t= m_t^2/{m_{H^-}^2}.
\eea
Using $\sflav$ we find that the total contribution to $C_{10}$ is
$C^{Z}_{10T }(H^{\pm})=$ $C_{10}^{Z}(H^{\pm})-C_{10}^{' \
  Z}(H^{\pm})\in (O(10^{-3}),O(10^{-2}))$, adding up little to
the total of $C_{10}$, including the SM contributions.
For our set of experimental values, we obtain that $C^{\rm{SM}}_{10}=-4.13\pm 0.05$. In the SM the current accuracy for this coefficient is better than $0.1\%$ level when allowing only the top-quark mass and and the strong coupling constant to deviate from their corresponding central value \cite{Bobeth:2013uxa}. In principle, then the contributions from  supersymmetric particles could be disentangled from the SM uncertainty.

\begin{figure}[htp]
\centering
\includegraphics[width=7cm]{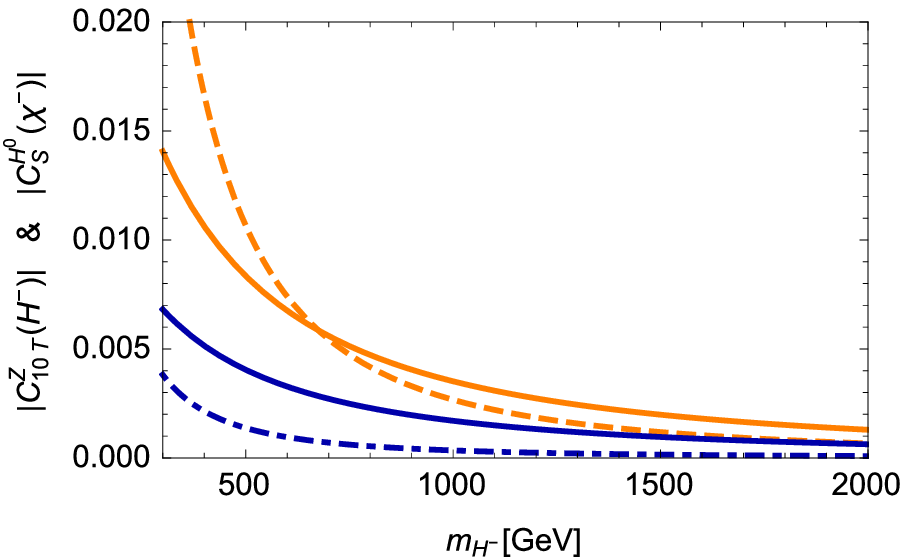}
\includegraphics[width=7cm]{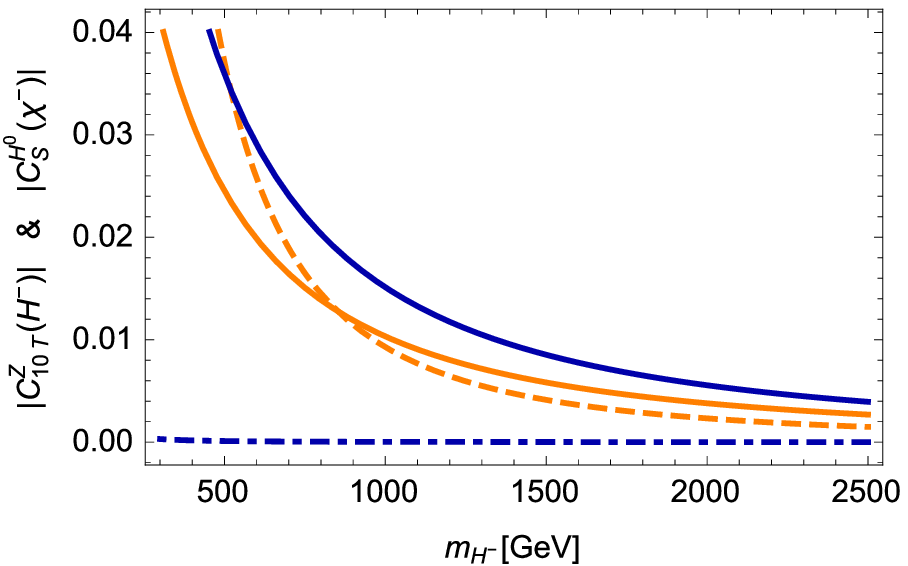}
\caption{\small{$|C_{10T}^Z(H^{\pm})|$ and $|C_S^{H^0}(\chi^{\pm})|$ as a function of $m_{H^\pm}(\approx m_{A})$
 for typical values of the pMSSM and the MSSM-30 samples. 
  The solid lines correspond to $|C_{10T}^Z(H^{\pm})|$ and the dashed
  and dot-dashed to  $|C_S^{H^0}(\chi^{\pm})|$.  For both plots,
  ($A_t$,$m_{\tilde t}$,$\mu$)=($1200, 1200, 100$) GeV.  For the left
  panel, the two values of $\tan\beta$ are $25$ (high) and $12$ (low),
  are plotted in orange-light- and dark-blue- respectively. These
  values correspond respectively to the typical values for the pMSSM
  and the MSSM-30 sample. For the right panel, we plotted the extreme
  values of the MSSM-30 sample, $\tan\beta=40,5$, plotted in
  orange-light- and dark-blue- respectively. In this last plot, we can
  appreciate the enhancement of 
$|C_{10T}^Z(H^{\pm})|$ for small values of $\tan\beta$. 
 \label{fig:CScomp_thLO} }}
\end{figure}

For illustration of our discussion, in \Figref{fig:CScomp_thLO} we
have made a comparison at LO of $C_{10T}^Z(H^{\pm})$ and
$C_S^{H^0}(\chi^{\pm})$ to emphasise the importance of the values of
$m_{H^{-}}$ and $\tan\beta$ in order to determine from which kind of diagram the
contributions to $\BRBsm$ are the most important 
 for each of the samples, the pMSSM and the MSSM-30. This comparison is made in the right panel of the figure, while in the left panel we show only cases which correspond to the MSSM-30 sample. 
Although in general for heavy spectra, both contributions are really small in comparison to the SM contributions, one can still appreciate the relevance of some supersymmetric particles. The solid lines correspond to $C_{10T}^Z(H^{\pm})$ and the dashed and dot-dashed to  $C_S^{H^0}(\chi^{\pm})$.  For both plots, ($A_t$,$m_{\tilde t}$,$\mu$)=($1200, 1200, 100$) GeV.  For the left panel the two values of $\tan\beta$ are $25$ (high) and $12$ (lower). These values correspond respectively to the typical values for the pMSSM and the MSSM-30 sample. For the right panel, we plotted the extreme values of the MSSM-30 sample, $\tan\beta=5, 40$.
 
 We can see that, as it is well established,  for values of $m_{H^-}$ below 1 TeV, the contribution from the large $\tan\beta$  values (here 25) coming from $C_S^{H^0}(\chi^{\pm})$ is the leading one (in \Figref{fig:CScomp_thLO} represented by the orange-light- dashed line). On the other hand,  contributions to both  $C_{10T}^Z(H^{\pm})$  and $C_S^{H^0}(\chi^{\pm})$ for values of $\tan\beta$ around 10 are typically less than a third of the corresponding values  when $\tan\beta>20$.  Since for the pMSSM sample, values for $\tan\beta$ above 20 dominate the sample, it is clear that the most important contributions come from $C_S^{H^0}(\chi^{\pm})$. For the MSSM-30 sample however, smaller values than 10 for  $\tan\beta$ are an important part of the sample and in this case they can give the largest supersymmetric contribution to the Wilson Coefficients via $C_{10T}^Z(H^{\pm})$. This is appreciated in the plot of the left in \Figref{fig:CScomp_thLO} where $C_{10T}^Z(H^{\pm})$ for $\tan\beta=5$ and $m_{H^-}>$ 700 GeV is the dominant of the supersymmetric contributions (solid blue-dark line).

 \subsection{Interplay of Wilson Coefficients}
In order to understand the different contributions to \eq{eq:thBR}, it
is customary to compare the relative size of the Wilson Coefficients
$C_S$ and $C_P$ to $C_{10}$. This is useful because $C_S$ and $C_P$ 
can only have supersymmetric contributions and supersymmetric
contributions  to $C_{10}$ are highly suppressed. 

In order to do this comparison, we compare the value of $C_{10}$ to
the Wilson Coefficients $C_{S}$ and $C_{P}$ but weighted in the same
way that $C_{10}$ contributes to $\BRBsm$, \eq{eq:thBR}. This allows a
direct comparison among  $C_{S}$, $C_{P}$ and $C_{10}$. The weighted
Wilson Coefficients $C_{S}$ and $C_{P}$ are respectively denoted by
$\hat{C}_{S}$ and $\hat{C}_{P}$: 
\bea
\hat{C}_{S} &\equiv&\frac{m_{B_s}^2}{2 m_{\mu} m_b}\sqrt{1-\frac{4 m_{\mu}^2}{m_{B_s}^2}} \ C_{S}\nonumber\\
\hat{C}_{P} &\equiv&\frac{m_{B_s}^2}{2 m_{\mu} m_b} \ C_{P}.
\eea
In \Figref{fig:WCM30comp}, we plot our results in planes $C_{10}$ vs $\hat{C}_{S}$ (we do not present $C_{10}$ vs $\hat{C}_{P}$ since  $\hat{C}_s\approx -\hat{C}_P$), comparing  the Wilson Coefficients for the two samples. For the pMSSM sample we present only the results when fixing the top mass value, since for our MSSM-30 sample the top mass was kept fixed. In this sense, even if the sample is not complete for the pMSSM, we give a fair comparison to the MSSM-30 sample and we can be sure that the increase (in absolute value) of the Wilson Coefficient $C_{10}$ arises due to the supersymmetric contributions.  While in the pMSSM, the center value of $C_{10}$ is -4.61 in the MSSM-30 is -4.64. Overall the contour plots for $C_{10}$ vs $C_S$ and $C_{10}$ vs $C_P$ cover a bigger area in the pMSSM (see also e.g. \cite{Arbey:2012ax}) than in the MSSM-30 but the one and two sigma regions of this last sample are shifted to the left, \Figref{fig:WCM30comp}.  As mentioned in Section \ref{sbsc:contsCPCS}, $C_{10}^{\rm{SM}}=-4.13\pm 0.05$ and the current accuracy is of the order $0.1\%$. Hence supersymmetric  contributions could be also disentangled from the SM error.

\begin{figure}[htp]
\centering
\includegraphics[width=8cm]{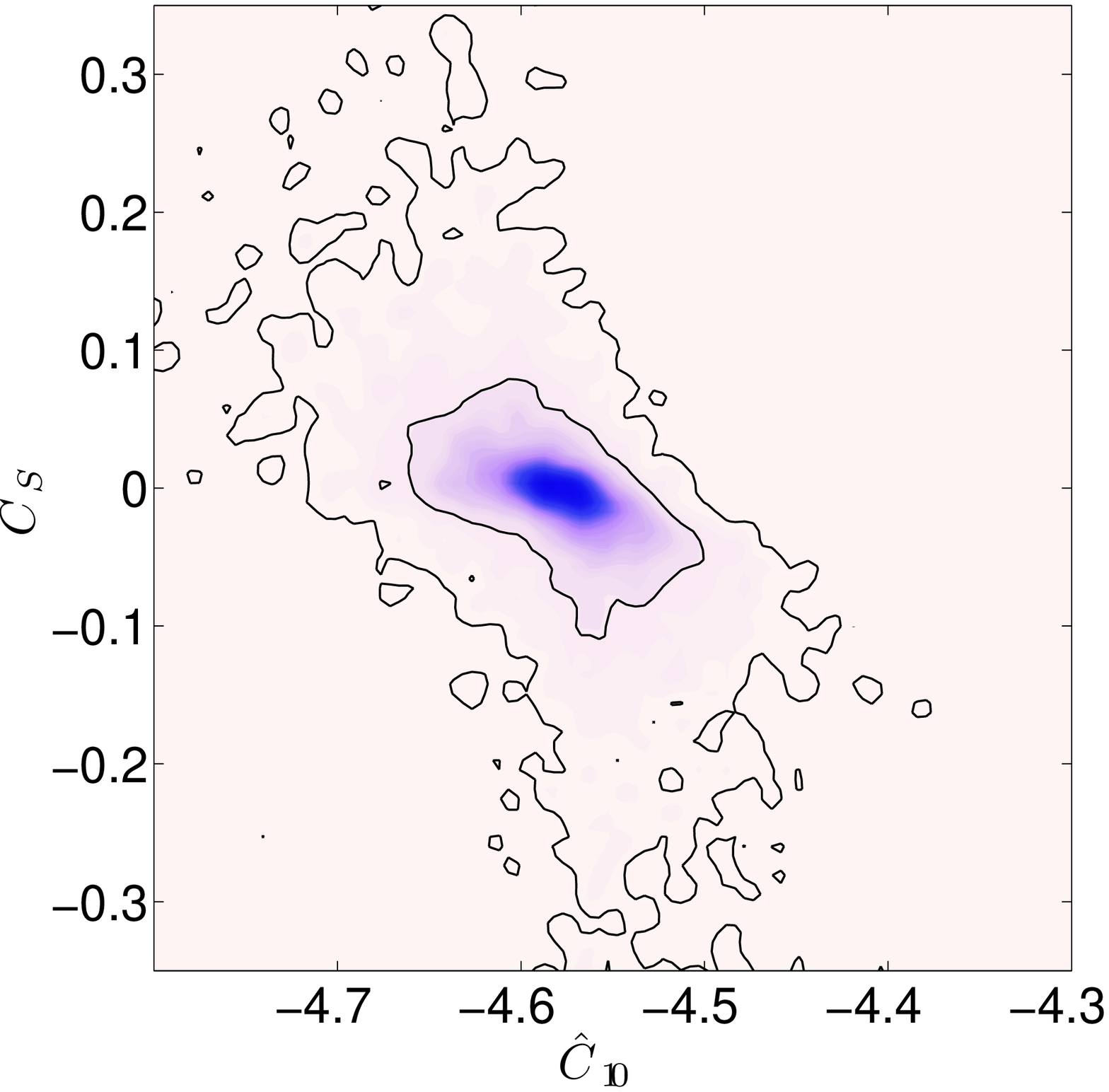} 
\includegraphics[width=8cm]{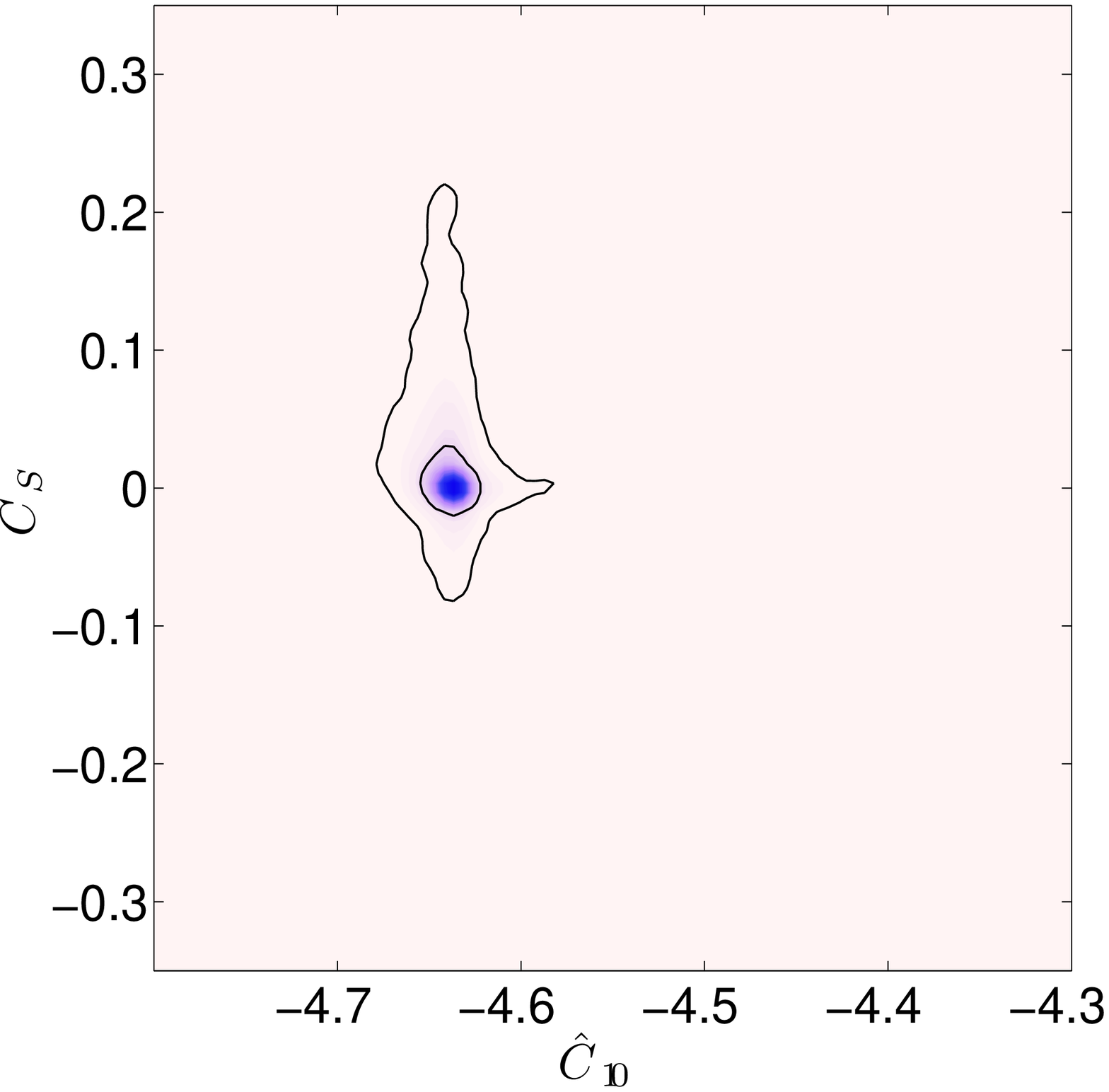} 
\caption{We compare here the Wilson Coefficients $C_{10}$, $C_S$ and  $C_P$ for the  pMSSM sample (left) and for the MSSM-30 sample (right). Within the pMSSM,  $C_{10} = -4.58 \pm 0.06$
  ranging from $-4.68$ to $-4.50$ at $95\%$ Bayesian probability
  region. For the MSSM-30, $C_{10} = -4.64 \pm 0.01$ ranging from
  $-4.66$ to $-4.63$ at $95\%$ Bayesian probability region.
\label{fig:WCM30comp}}
\end{figure}

From \Figref{fig:CScomp_thLO} we can see that for large values of $m_{H^{-}}$ ($>$1500 GeV)  the contributions from $|C_{10T}^Z(H^{\pm})|$ and $|C_S^{H^0}(\chi^{\pm})$ become quite similar, especially for lower values of $\tan\beta$ (approximately below 20). On the other hand, since for the pMSSM sample, values for $\tan\beta$ above 20 dominate the sample,  the most important contributions come from $C_S^{H^0}(\chi^{\pm})$. For the MSSM-30 sample however, smaller values than 10 for  $\tan\beta$ are an important part of the sample and in this case they can give the largest supersymmetric contribution to the Wilson Coefficients via $C_{10T}^Z(H^{\pm})$, as mentioned before (\Figref{fig:CScomp_thLO}).

We recall the reader that the present work builds further on the project for  MSSM explorations within systematically built frameworks, in this case a specific frame for flavor violation and a Bayesian approach for deriving inference from experimental data. The specific flavor violation determined  by the MSSM-30 is a realistic one, owning to the fact that it can be understood as taking into account the running of off-diagonal elements of soft-squared masses and trilinears. In contrast, the pMSSM sets these elements to zero to start with. Since there are many supersymmetric contributions in a less constrained MSSM, it is natural to expect that these contributions will have a constructive interference instead of a destructive one, resulting in the increase of  the value of $\Bsm$. This is effectively what is happening in the MSSM-30 in comparison to the pMSSM: owning to the fact that the MSSM-30 is less constrained, it does increase the value of  $\Bsm$ in comparison to the pMSSM. It is out of the scope of this work to identify a region where actually cancellations could take place, but it is definitely a project that it should be peformed.

 \subsection{Interplay of off-diagonal soft-squared elements}
 It is customary to assess the impact of flavor violation in terms of the flavor violating parameters 
 \bea
 \delta_{Q XY}^{ij}= \frac{\left(\widehat{M}^2_Q\right)^{ij}_{XY}}{\sqrt{    \left(\widehat{M}^2_Q\right)^{ii}_{XX}  \left(\widehat{M}^2_Q\right)^{jj}_{YY}   }},
 \eea
 which measure the amount of off-diagonal allowed contributions 
 constrained by all relevant flavor observables (see Table 1 of
 \cite{AbdusSalam:2014uea}). 
 Here  $(\widehat{M}^2_Q)_{LR}=-A_D v_D + \mu\tan\beta m_D$ ($v_D=v_U/\tan\beta$ and $m_D$ the diagonal mass matrix of D quarks), $(\widehat{M}^2_Q)_{RR}=M^2_D$,  $(\widehat{M}^2_Q)_{LL}=M^2_Q$, the mass matrices without hat are those appearing in \eq{mfvpar30}.  Here we present only the relevant parameters for $\BRBsm$. 
   In order to make manifest how the Z penguin contributions dominate the supersymmetric contributions of the coefficient $C_{10}$ for most part of the parameter space. We present in \Figref{fig:deltaLR_BR} (top row) the individual contributions to $\BRBsm$ from Higgs penguin, Z penguin and Box diagrams, as a function of $\left(\delta_{QLR}\right)^{23}$.
 \begin{figure}[htp]
   \includegraphics[width=5.2cm]{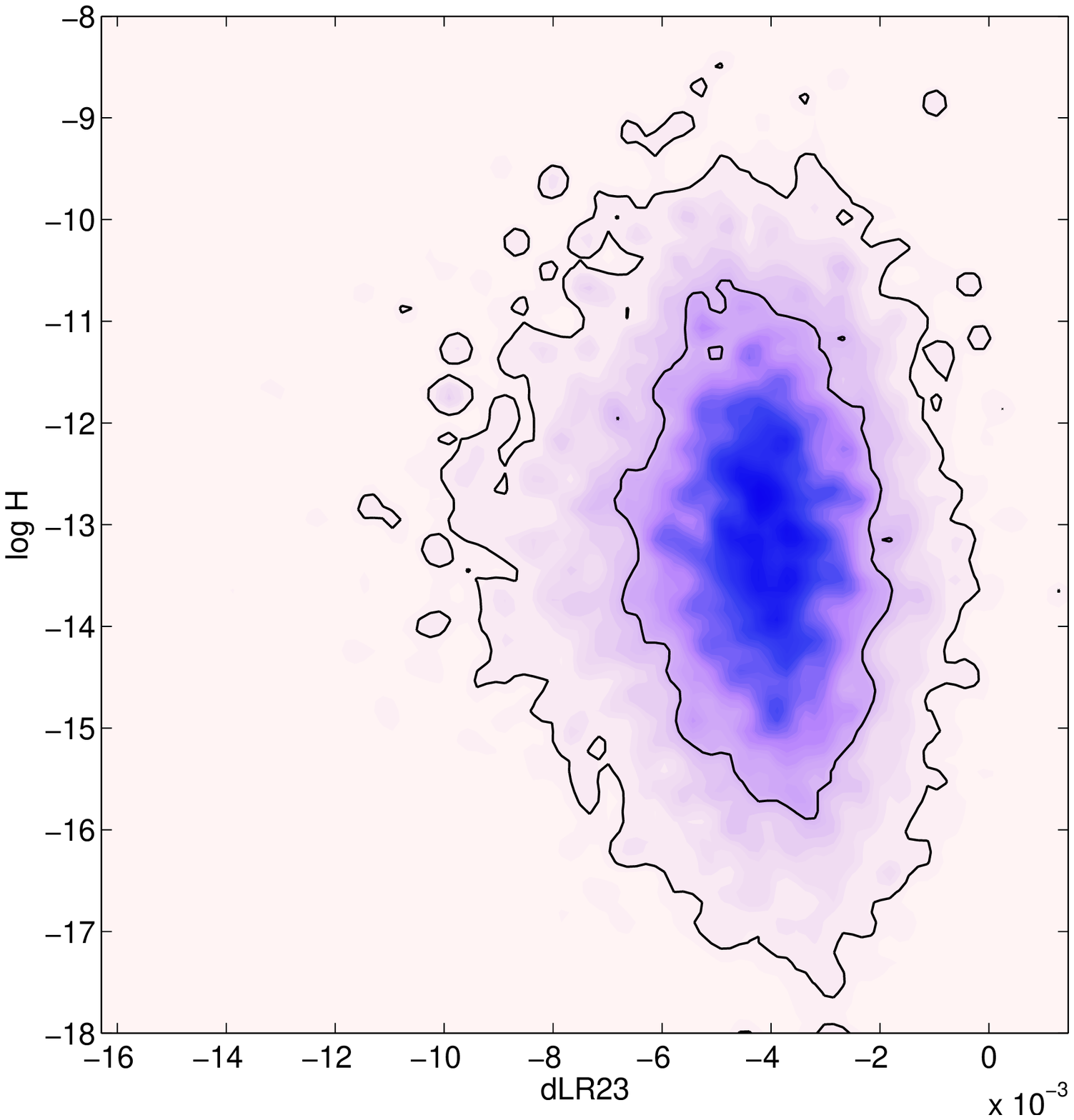}
   \includegraphics[width=5.2cm]{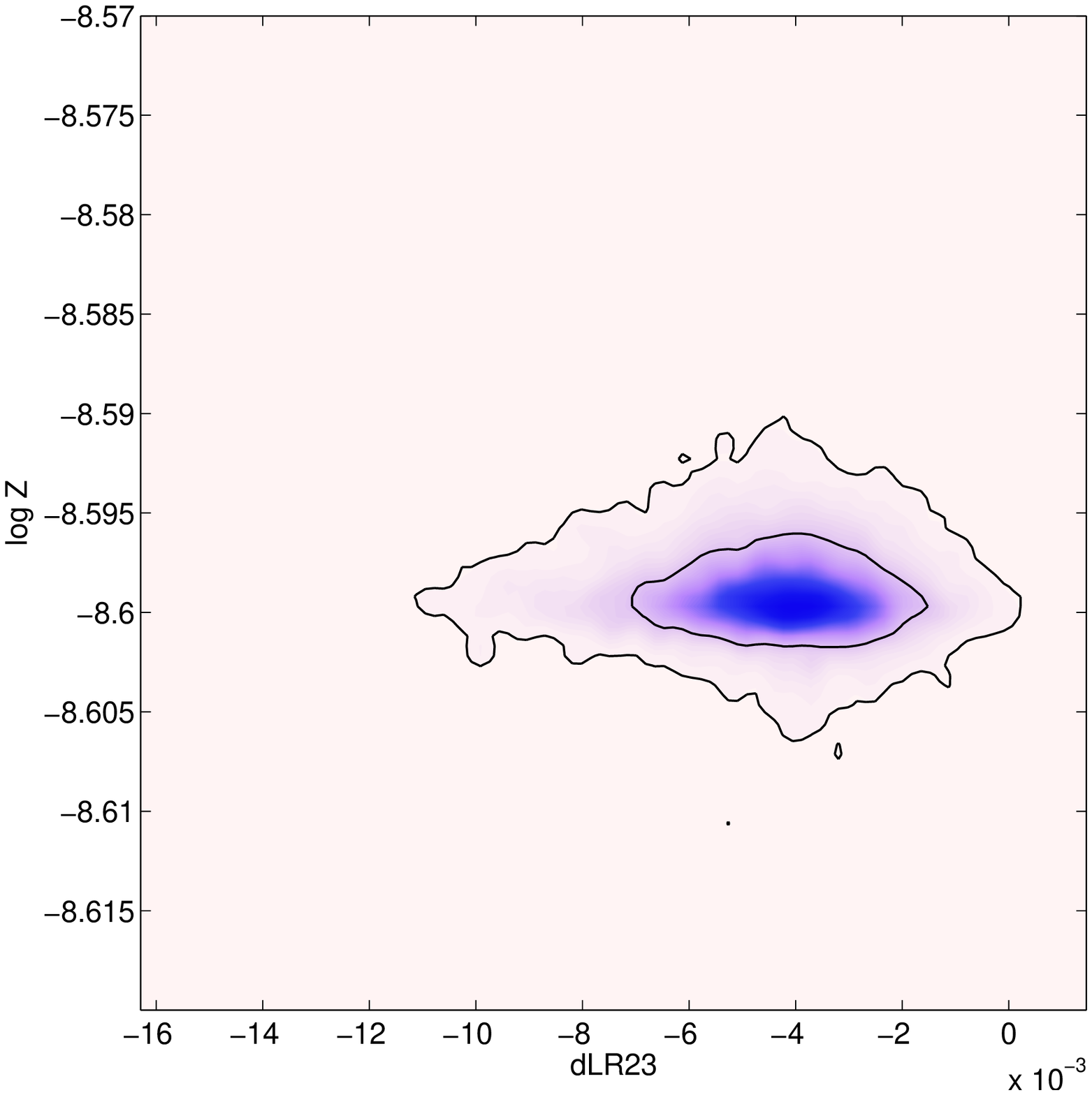}
   \includegraphics[width=5.2cm]{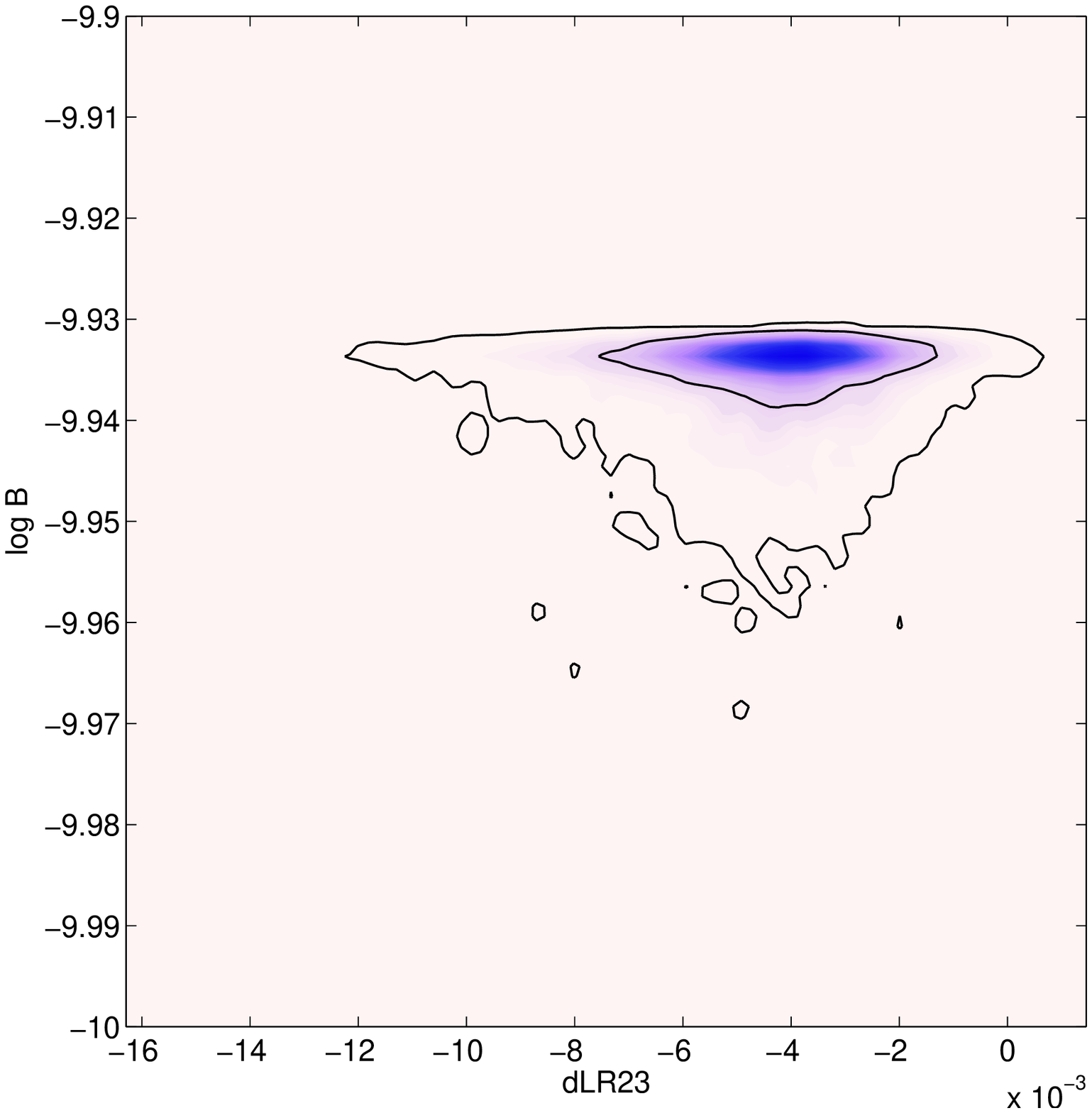}\\
   \includegraphics[width=5.2cm]{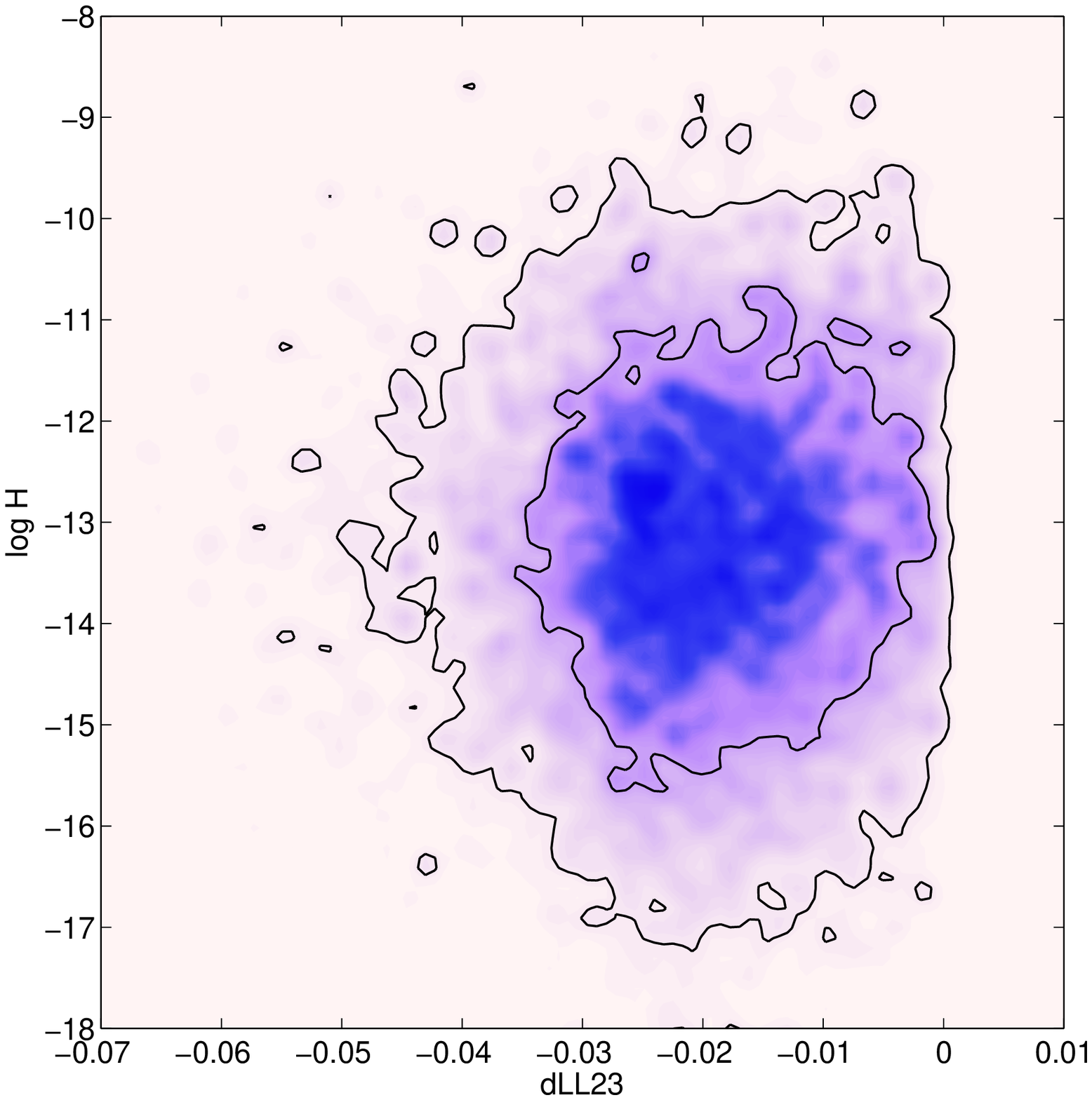}
   \includegraphics[width=5.2cm]{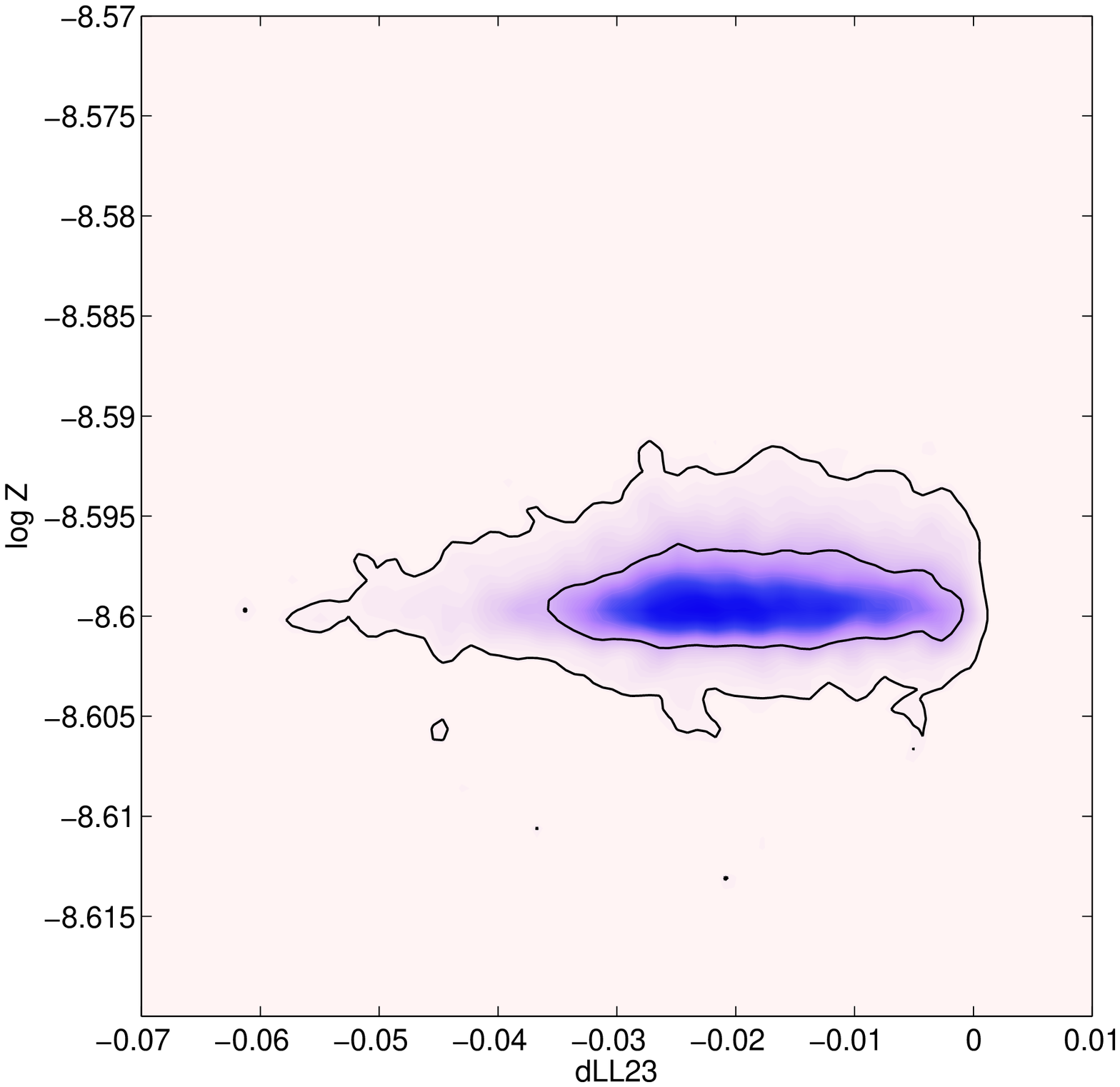}
   \includegraphics[width=5.2cm]{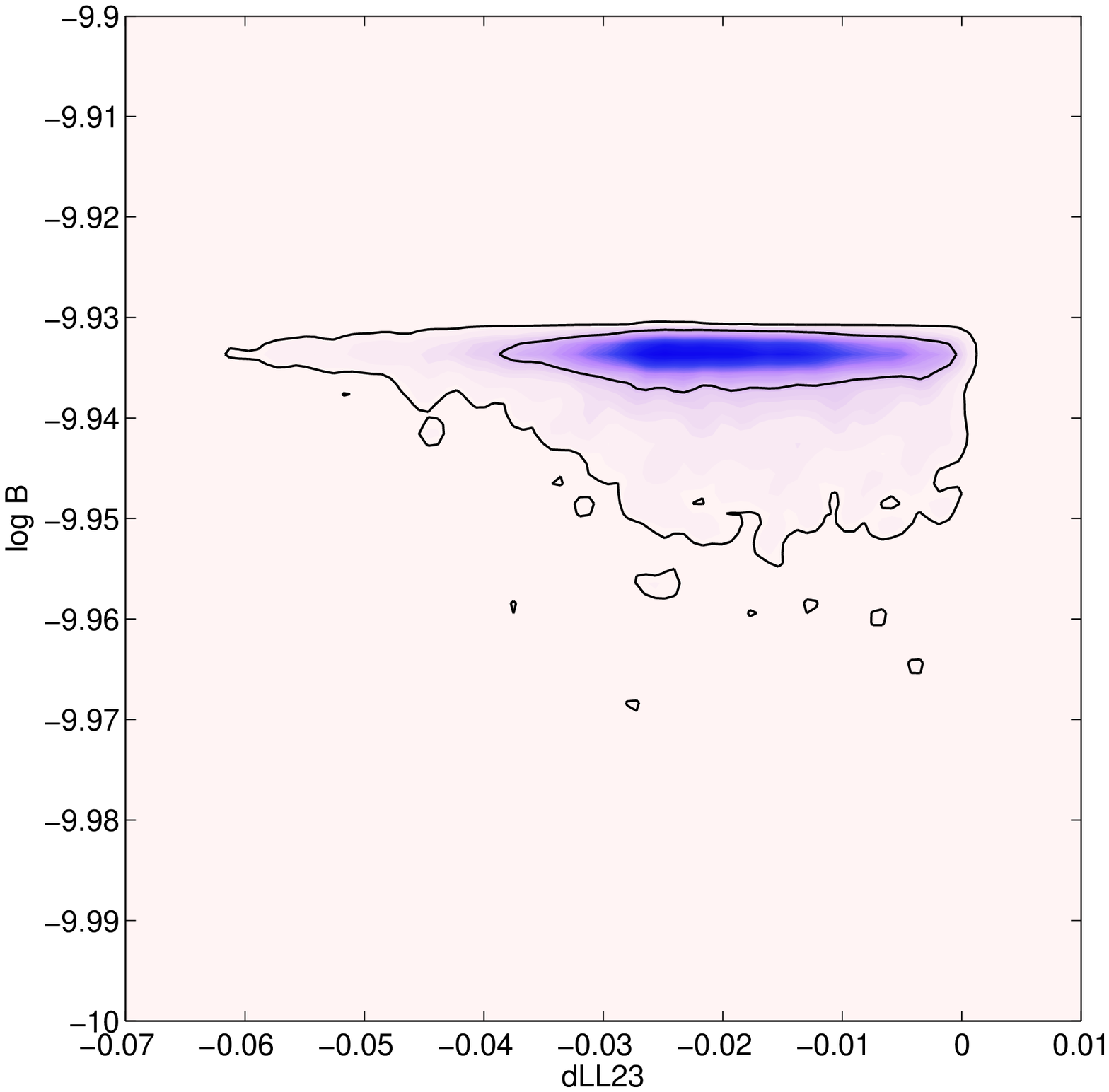}
   \caption{Contributions to $\BRBsm$ from the parameters
     ${\rm dLR23} = \left(\delta_{QLR}\right)^{23}$ and
     ${\rm dLL23} = \left(\delta_{QLL}\right)^{23}$ (top and bottom rows
     respectively).  $\rm{Log}$ H, $\rm{Log}$ Z and $\rm{Log}$ B
     represent respectively the logarithms of the absolute values of
     Higgs Penguin, Z penguin and Box contributions to the total
     $\BRBsm$. The outer and inner contours enclose the $95\%$ and
     $68\%$ Bayesian probability regions
     respectively. \label{fig:deltaLR_BR}}   
 \end{figure}
In this figure, we can clearly appreciate how Z penguin contributions are in general the most dominant for practically all values of $\left(\delta_{QLR}\right)^{23}$ and as emphasized in \cite{Chankowski:2000ng}, for small values of $\tan\beta$ ($\leq$ 20),  Higgs penguin contributions are small. Box contributions are also small, bu they tend to be even bigger than the Higgs penguin contributions.
In the second row of  \Figref{fig:deltaLR_BR} we represent the individual contributions to $\BRBsm$ from Higgs penguin, Z penguin and Box diagrams, as a function of $\left(\delta_{QLL}\right)^{23}$.  As it is usual with LR flavor violation, it tends to be more constrained, as in this case it is allowed to be only $O(10^{-3})$, while LL flavor violation it can be of $O(10^{-2})$.

\section{Conclusions}
\label{sec:summary}
We have continued with our studies in order to explore features of the
MSSM  by using Bayesian statistical
techniques on systematically constructed, symmetry-guided, MSSM
  frameworks beyond the traditional constructions. Here, the
phenomenological framework considered is the 30-parameter-MSSM, called
MSSM-30, and the observable of interest is the $\Bsm$ decay. The
 measured branching ratio $\BRBsm$
is compatible with the SM prediction but it still has a large (order $20\%$)
uncertainty. Future precision measurements of this observable will be
excellent for assessing the MSSM as New Physics beyond the SM. Within
the MSSM-30 a posterior sample was considered,  for
  which a $15\%$ uncertainty on the measured $\BRBsm$ would
favour  a sub-TeV pseudo-scalar Higgs boson. 

Knowing that the decay $\Bsm$ is a good indicator for assessing
  models with extended Higgs sectors, we compared the MSSM-30 to the
  pMSSM to see if there are any physics or features in the parameters of 
  the former which are not accessible in the latter. It turned out to
  be the case since the MSSM-30 sample prefers  lower values of
  $\tan\beta$ $\sim (10,20)$ in comparison to the pMSSM which prefers
  $\tan\beta \sim (20,40)$. It is then possible to find higher values
  for $\BRBsm$  in the MSSM-30 mainly due to bigger contributions
  coming from diagrams involving charginos and Z-penguin
  diagrams. 
  
We found that  the best way to analyse the contributions to the
branching ratio $\BRBsm$ was by comparing kind of diagrams: Z
penguin, box and Higgs penguin diagrams, instead of comparing contributions
of supersymmetric particles (i.e. gluinos, neutralinos or charginos).
The reason is that in the pMSSM the supersymmetric
contributions become quite suppressed due to the large values of
$m_A$ and  $m_{\tilde t}$ (well into the multi-TeV region) 
and that in the SM the $Z$ penguin and box contributions are dominant
( $\sim 75\%$ and $24\%$  respectively). This last fact then in
principle helps to look for contributions coming from
supersymmetry.  In the pMSSM  it is well established that the major BSM contributions to $\BRBsm$  come from a Higgs penguin and a chargino in the loop. Hence, this is a specific example of how analysis by kind of diagrams becomes relevant.

 The MSSM-30 has by  construction  non-zero off-diagonal soft-squared mass terms, contrary to the pMSSM where they are set to zero by hand.  When analysing both samples using a Bayesian fit, both samples are constrained by the same observables. Therefore, if no cancellations appear, the value of the  allowed effective off-diagonal soft-squared mass terms should be of the same order in both samples.  The best place to look for the difference between both samples is in the contribution coming from the charginos, mainly from Higgs-penguins, (\Figref{fig:ParticlesContr}) where we can see that the distributions from pMSSM and MSSM-30 are different but  mostly indistinguishable towards higher values of $\BRBsm$.

In order to  assess the impact of BSM contributions to B observables, it is customary to compare
the size of the Wilson Coefficients which receive the majority of the
BSM contributions with the Wilson Coefficients present only in the SM
case. For the decay $\Bsm$, the relevant coefficients are $C_S$ and
$C_P$,  which are scalar operators sensitive to the chirality of 
  BSM contributions. In the SM the only contribution comes
from the vector operator $C^{SM}_{10}$, for which we find a value of
$-4.13 \pm 0.05$. We have compared the coefficients 
$C_S$ and $C_P$ to $C_{10}$, using $C_{10}$ vs $C_{S}$ and  $C_{10}$
vs $C_{P}$ planes and found that in the MSSM-30, $C_S$ and $C_P$
represent typically only a O(1\%) contribution to the branching ratio
$\Bsm$.
Within the pMSSM  $C_{10} = -4.58 \pm 0.06$
  ranging from $-4.68$ to $-4.50$ at $95\%$ Bayesian probability
  region. For the MSSM-30, $C_{10} = -4.64 \pm 0.01$ ranging from
  $-4.66$ to $-4.63$ at $95\%$ Bayesian probability region. The
  MSSM-30 is more severely away from the SM value compared to the
  pMSSM. 
The current SM
  accuracy in determining the value of  $C_{10}$ is of the order
  $0.1\%$ and therefore supersymmetric contributions can be
  disentangled from the SM uncertainty.

Finally, as an outlook we highlight that future improvements of
the measurement 
of $\BRBsm$, along with the measurement of other observables, like
$\ADF$ and $\BRBdm$ which are correlated with   $\BRBsm$, will play a
crucial role in shaping the parameter space of the MSSM-30. 
 On the other hand, systematically constructed frameworks
which can capture the flavor structure of the MSSM, like the one
presented here, should be favored over simplified scenarios which
cannot capture the rich flavor structure of the MSSM.

\section*{Acknowledgments}
We would like to thank Janusz Rosiek for help and comments regarding
$\rm{SUSY\_FLAVOR}$. L.VS thanks R. Fleischer for comments regarding
$\BRBsm$ and the Abdus Salam International Center for Theoretical
Physics, Italy, for support and hospitality during the last stages of
this project. During an earlier stage of this work, S.S.A. was funded
at INFN, Sezione di Roma, under the European Research Council's
Seventh Framework Programme (FP/2007-2013)/ERC Grant Agreement
no. 279972, NPflavor. 



\begin{thebibliography}{}

\bibitem{Bobeth:2001sq}
  C.~Bobeth, T.~Ewerth, F.~Kruger and J.~Urban,
  Phys.\ Rev.\ D {\bf 64} (2001) 074014
  doi:10.1103/PhysRevD.64.074014
  [hep-ph/0104284].



\bibitem{Bobeth:2002ch}
  C.~Bobeth, T.~Ewerth, F.~Kruger and J.~Urban,
  Phys.\ Rev.\ D {\bf 66} (2002) 074021
  doi:10.1103/PhysRevD.66.074021
  [hep-ph/0204225].



\bibitem{Huang:1998vb}
  C.~S.~Huang, W.~Liao and Q.~S.~Yan,
  Phys.\ Rev.\ D {\bf 59} (1999) 011701
  doi:10.1103/PhysRevD.59.011701
  [hep-ph/9803460].



\bibitem{Hamzaoui:1998nu}
  C.~Hamzaoui, M.~Pospelov and M.~Toharia,
  Phys.\ Rev.\ D {\bf 59} (1999) 095005
  doi:10.1103/PhysRevD.59.095005
  [hep-ph/9807350].



\bibitem{Choudhury:1998ze}
  S.~R.~Choudhury and N.~Gaur,
  Phys.\ Lett.\ B {\bf 451} (1999) 86
  doi:10.1016/S0370-2693(99)00203-8
  [hep-ph/9810307].



\bibitem{Babu:1999hn}
  K.~S.~Babu and C.~F.~Kolda,
  Phys.\ Rev.\ Lett.\  {\bf 84} (2000) 228
  doi:10.1103/PhysRevLett.84.228
  [hep-ph/9909476].



\bibitem{Huang:2000sm}
  C.~S.~Huang, W.~Liao, Q.~S.~Yan and S.~H.~Zhu,
  Phys.\ Rev.\ D {\bf 63} (2001) 114021
  Erratum: [Phys.\ Rev.\ D {\bf 64} (2001) 059902]
  doi:10.1103/PhysRevD.64.059902, 10.1103/PhysRevD.63.114021
  [hep-ph/0006250].



\bibitem{AbdusSalam:2009qd}
  S.~S.~AbdusSalam, B.~C.~Allanach, F.~Quevedo, F.~Feroz and M.~Hobson,
  Phys.\ Rev.\ D {\bf 81} (2010) 095012
  doi:10.1103/PhysRevD.81.095012
  [arXiv:0904.2548 [hep-ph]].



\bibitem{Arbey:2012ax}
  A.~Arbey, M.~Battaglia, F.~Mahmoudi and D.~Martínez Santos,
  Phys.\ Rev.\ D {\bf 87} (2013) no.3,  035026
  doi:10.1103/PhysRevD.87.035026
  [arXiv:1212.4887 [hep-ph]].



\bibitem{Altmannshofer:2017wqy}
  W.~Altmannshofer, C.~Niehoff and D.~M.~Straub,
  JHEP {\bf 1705} (2017) 076
  doi:10.1007/JHEP05(2017)076
  [arXiv:1702.05498 [hep-ph]].



\bibitem{Aaltonen:2011fi}
  T.~Aaltonen {\it et al.} [CDF Collaboration],
  Phys.\ Rev.\ Lett.\  {\bf 107} (2011) 191801
  [Phys.\ Rev.\ Lett.\  {\bf 107} (2011) 239903]
  Addendum: [Phys.\ Rev.\ Lett.\  {\bf 107} (2011) no.23,  239903]
  doi:10.1103/PhysRevLett.107.191801, 10.1103/PhysRevLett.107.239903
  [arXiv:1107.2304 [hep-ex]].



\bibitem{Abazov:2010fs}
  V.~M.~Abazov {\it et al.} [D0 Collaboration],
  Phys.\ Lett.\ B {\bf 693} (2010) 539
  doi:10.1016/j.physletb.2010.09.024
  [arXiv:1006.3469 [hep-ex]].



\bibitem{Aaij:2012ac}
  R.~Aaij {\it et al.} [LHCb Collaboration],
  Phys.\ Rev.\ Lett.\  {\bf 108} (2012) 231801
  doi:10.1103/PhysRevLett.108.231801
  [arXiv:1203.4493 [hep-ex]].



\bibitem{Chatrchyan:2012rga}
  S.~Chatrchyan {\it et al.} [CMS Collaboration],
  JHEP {\bf 1204} (2012) 033
  doi:10.1007/JHEP04(2012)033
  [arXiv:1203.3976 [hep-ex]].



\bibitem{Aad:2012pn}
  G.~Aad {\it et al.} [ATLAS Collaboration],
  Phys.\ Lett.\ B {\bf 713} (2012) 387
  doi:10.1016/j.physletb.2012.06.013
  [arXiv:1204.0735 [hep-ex]].



\bibitem{Aaij:2012nna}
  R.~Aaij {\it et al.} [LHCb Collaboration],
  Phys.\ Rev.\ Lett.\  {\bf 110} (2013) no.2,  021801
  doi:10.1103/PhysRevLett.110.021801
  [arXiv:1211.2674 [hep-ex]].



\bibitem{CMS:2014xfa}
  V.~Khachatryan {\it et al.} [CMS and LHCb Collaborations],
  Nature {\bf 522} (2015) 68
  doi:10.1038/nature14474
  [arXiv:1411.4413 [hep-ex]].



\bibitem{Mulder:2017hug}
  M.~Mulder [LHCb Collaboration],
  arXiv:1705.03274 [hep-ex].
\bibitem{Aaij:2017vad} 
  R.~Aaij {\it et al.} [LHCb Collaboration],
  Phys.\ Rev.\ Lett.\  {\bf 118}, no. 19, 191801 (2017)
  doi:10.1103/PhysRevLett.118.191801
  [arXiv:1703.05747 [hep-ex]].
  



\bibitem{Colangelo:2008qp}
  G.~Colangelo, E.~Nikolidakis and C.~Smith,
  Eur.\ Phys.\ J.\ C {\bf 59} (2009) 75
  doi:10.1140/epjc/s10052-008-0796-y
  [arXiv:0807.0801 [hep-ph]].



\bibitem{AbdusSalam:2014uea}
  S.~S.~AbdusSalam, C.~P.~Burgess and F.~Quevedo,
  JHEP {\bf 1502} (2015) 073
  doi:10.1007/JHEP02(2015)073
  [arXiv:1411.1663 [hep-ph]].



\bibitem{Dedes:2001fv}
  A.~Dedes, H.~K.~Dreiner and U.~Nierste,
  Phys.\ Rev.\ Lett.\  {\bf 87} (2001) 251804
  doi:10.1103/PhysRevLett.87.251804
  [hep-ph/0108037].



\bibitem{Ellis:2005sc}
  J.~R.~Ellis, K.~A.~Olive and V.~C.~Spanos,
  Phys.\ Lett.\ B {\bf 624} (2005) 47
  doi:10.1016/j.physletb.2005.07.066
  [hep-ph/0504196].



\bibitem{Heinemeyer:2008fb}
  S.~Heinemeyer, X.~Miao, S.~Su and G.~Weiglein,
  JHEP {\bf 0808} (2008) 087
  doi:10.1088/1126-6708/2008/08/087
  [arXiv:0805.2359 [hep-ph]].



\bibitem{Alok:2009wk}
  A.~K.~Alok and S.~K.~Gupta,
  Eur.\ Phys.\ J.\ C {\bf 65} (2010) 491
  doi:10.1140/epjc/s10052-009-1163-3
  [arXiv:0904.1878 [hep-ph]].



\bibitem{Mahmoudi:2012un}
  F.~Mahmoudi, S.~Neshatpour and J.~Orloff,
  JHEP {\bf 1208} (2012) 092
  doi:10.1007/JHEP08(2012)092
  [arXiv:1205.1845 [hep-ph]].



\bibitem{Buchmueller:2012hv}
  O.~Buchmueller {\it et al.},
  Eur.\ Phys.\ J.\ C {\bf 72} (2012) 2243
  doi:10.1140/epjc/s10052-012-2243-3
  [arXiv:1207.7315 [hep-ph]].



\bibitem{Workgroup:2017myk}
  F.~U.~Bernlochner {\it et al.} [GAMBIT Collaboration],
  arXiv:1705.07933 [hep-ph].



\bibitem{Djouadi:1998di}
  A.~Djouadi {\it et al.} [MSSM Working Group],
  hep-ph/9901246.


\bibitem{Kadota:2011cr} 
  K.~Kadota, G.~Kane, J.~Kersten and L.~Velasco-Sevilla,
  Eur.\ Phys.\ J.\ C {\bf 72}, 2004 (2012)
  doi:10.1140/epjc/s10052-012-2004-3
  [arXiv:1107.3105 [hep-ph]].


\bibitem{Ellis:2016qra} 
  J.~Ellis, K.~Olive and L.~Velasco-Sevilla,
  Eur.\ Phys.\ J.\ C {\bf 76}, no. 10, 562 (2016)
  doi:10.1140/epjc/s10052-016-4398-9
  [arXiv:1605.01398 [hep-ph]].
  
\bibitem{Poh:2015wta} 
  Z.~Poh and S.~Raby,
  Phys.\ Rev.\ D {\bf 92}, no. 1, 015017 (2015)
  doi:10.1103/PhysRevD.92.015017
  [arXiv:1505.00264 [hep-ph]]. 
  

\bibitem{AbdusSalam:2008uv}
  S.~S.~AbdusSalam,
  AIP Conf.\ Proc.\  {\bf 1078} (2009) 297
  doi:10.1063/1.3051939
  [arXiv:0809.0284 [hep-ph]].



\bibitem{AbdusSalam:2010qp}
  S.~S.~AbdusSalam and F.~Quevedo,
  Phys.\ Lett.\ B {\bf 700} (2011) 343
  doi:10.1016/j.physletb.2011.02.065
  [arXiv:1009.4308 [hep-ph]].



\bibitem{AbdusSalam:2011hd}
  S.~S.~AbdusSalam,
  Phys.\ Lett.\ B {\bf 705} (2011) 331
  doi:10.1016/j.physletb.2011.10.023
  [arXiv:1106.2317 [hep-ph]].



\bibitem{AbdusSalam:2011fc}
  S.~S.~AbdusSalam {\it et al.},
  Eur.\ Phys.\ J.\ C {\bf 71} (2011) 1835
  doi:10.1140/epjc/s10052-011-1835-7
  [arXiv:1109.3859 [hep-ph]].



\bibitem{AbdusSalam:2012sy}
  S.~S.~AbdusSalam and D.~Choudhury,
  Universal J.\ Phys.\ Appl.\  {\bf 2} (2014) no.3,  155
  doi:10.13189/ujpa.2014.020303
  [arXiv:1210.3331 [hep-ph]].



\bibitem{AbdusSalam:2012ir}
  S.~S.~AbdusSalam,
  Phys.\ Rev.\ D {\bf 87} (2013) no.11,  115012
  doi:10.1103/PhysRevD.87.115012
  [arXiv:1211.0999 [hep-ph]].



\bibitem{AbdusSalam:2013qba}
  S.~S.~AbdusSalam,
  Int.\ J.\ Mod.\ Phys.\ A {\bf 29} (2014) no.27,  1450160
  doi:10.1142/S0217751X14501607
  [arXiv:1312.7830 [hep-ph]].



\bibitem{AbdusSalam:2015uba}
  S.~S.~AbdusSalam and L.~Velasco-Sevilla,
  Phys.\ Rev.\ D {\bf 94} (2016) no.3,  035026
  doi:10.1103/PhysRevD.94.035026
  [arXiv:1506.02499 [hep-ph]].


\bibitem{Belanger:2008sj}
  G.~Belanger, F.~Boudjema, A.~Pukhov and A.~Semenov,
  Comput.\ Phys.\ Commun.\  {\bf 180} (2009) 747
  [arXiv:0803.2360 [hep-ph]].

\bibitem{Skands:2003cj}
  P.~Z.~Skands, B.~C.~Allanach, H.~Baer, C.~Balazs, G.~Belanger, F.~Boudjema, A.~Djouadi and R.~Godbole {\it et al.},
  JHEP {\bf 0407} (2004) 036
  [hep-ph/0311123].

\bibitem{Allanach:2008qq}
  B.~C.~Allanach, C.~Balazs, G.~Belanger, M.~Bernhardt, F.~Boudjema, D.~Choudhury, K.~Desch and U.~Ellwanger {\it et al.},
  Comput.\ Phys.\ Commun.\  {\bf 180} (2009) 8
  [arXiv:0801.0045 [hep-ph]].

\bibitem{Porod:2011nf}
  W.~Porod and F.~Staub,
  Comput.\ Phys.\ Commun.\  {\bf 183} (2012) 2458
  [arXiv:1104.1573 [hep-ph]].


\bibitem{Crivellin:2012jv}
  A.~Crivellin, J.~Rosiek, P.~H.~Chankowski, A.~Dedes, S.~Jaeger and P.~Tanedo,
  Comput.\ Phys.\ Commun.\  {\bf 184} (2013) 1004
  doi:10.1016/j.cpc.2012.11.007
  [arXiv:1203.5023 [hep-ph]].



\bibitem{Chankowski:2000ng}
  P.~H.~Chankowski and L.~Slawianowska,
  Phys.\ Rev.\ D {\bf 63} (2001) 054012
  doi:10.1103/PhysRevD.63.054012
  [hep-ph/0008046].



\bibitem{Logan:2000iv}
  H.~E.~Logan and U.~Nierste,
  Nucl.\ Phys.\ B {\bf 586} (2000) 39
  doi:10.1016/S0550-3213(00)00417-X
  [hep-ph/0004139].




\bibitem{Dedes:2008iw}
  A.~Dedes, J.~Rosiek and P.~Tanedo,
  Phys.\ Rev.\ D {\bf 79} (2009) 055006
  doi:10.1103/PhysRevD.79.055006
  [arXiv:0812.4320 [hep-ph]].


\bibitem{verzo}
M. Verzocchi, talk at the 34th International Conference on High Energy
Physics (ICHEP 2008): Philadelphia, Pennsylvania (unpublished).


\bibitem{:2005ema}
  {\bf ALEPH} Collaboration,
  Phys. Rept. {\bf 427} (2006) 257.

\bibitem{Aaij:2011rja}
  R.~Aaij {\it et al.}  [LHCb Collaboration],
  Phys.\ Lett.\ B {\bf 699} (2011) 330
  [arXiv:1103.2465 [hep-ex]].

\bibitem{Abulencia:2006ze}
  A.~Abulencia {\it et al.}  [CDF Collaboration],
  Phys.\ Rev.\ Lett.\  {\bf 97} (2006) 242003.

\bibitem{Aubert:2004kz}
  B.~Aubert {\it et al.}  [BaBar Collaboration],
  Phys.\ Rev.\ Lett.\  {\bf 95} (2005) 041804.

\bibitem{Barberio:2008fa}
  E.~Barberio {\it et al.}  [Heavy Flavor Averaging Group
    Collaboration],
  arXiv:0808.1297 [hep-ex].

\bibitem{0803.0547}
  E.~Komatsu {\it et al.}  [WMAP Collaboration],
  Astrophys.\ J.\ Suppl.\  {\bf 180} (2009) 330.

\bibitem{ATLAS:2013mma}
  [ATLAS Collaboration],
  ATLAS-CONF-2013-014.


\bibitem{CMS:yva}
  [CMS Collaboration],
  CMS-PAS-HIG-13-005.

\bibitem{McNabb:2004tj}
  R.~McNabb [Muon g-2 Collaboration],
  hep-ex/0407008.

\bibitem{Barberio:2007cr}
  E.~Barberio {\it et al.}  [Heavy Flavor Averaging Group (HFAG) Collaboration],
  arXiv:0704.3575 [hep-ex].

\bibitem{Nakamura:2010zzi}
  K.~Nakamura {\it et al.}  [Particle Data Group Collaboration],
  J.\ Phys.\ G {\bf 37} (2010) 075021.

\bibitem{Regan:2002ta}
  B.~C.~Regan, E.~D.~Commins, C.~J.~Schmidt and D.~DeMille,
  Phys.\ Rev.\ Lett.\  {\bf 88} (2002) 071805.

\bibitem{Porod:2003um}
  W.~Porod,
  Comput.\ Phys.\ Commun.\  {\bf 153} (2003) 275
  [hep-ph/0301101].


\bibitem{Heinemeyer:2006px}
  S.~Heinemeyer, W.~Hollik, D.~Stockinger, A.~M.~Weber and G.~Weiglein,
  JHEP {\bf 0608} (2006) 052
  [hep-ph/0604147].

\bibitem{Heinemeyer:2007bw}
  S.~Heinemeyer, W.~Hollik, A.~M.~Weber and G.~Weiglein,
  JHEP {\bf 0804} (2008) 039
  [arXiv:0710.2972 [hep-ph]].

\bibitem{Feroz:2007kg}
  F.~Feroz and M.~P.~Hobson,
  Mon.\ Not.\ Roy.\ Astron.\ Soc.\  {\bf 384} (2008) 449
  [arXiv:0704.3704].

\bibitem{Feroz:2008xx}
  F.~Feroz, M.~P.~Hobson and M.~Bridges,
  Mon.\ Not.\ Roy.\ Astron.\ Soc.\  {\bf 398} (2009) 1601
  [arXiv:0809.3437].

\bibitem{Skilling}
  J.~{Skilling}, 
  in {American Institute of Physics Conference Series}
  (R.~{Fischer}, R.~{Preuss}, and U.~V. {Toussaint}, eds.),
  pp.~395--405, Nov., 2004.

\bibitem{Pospelov:2005pr}
  M.~Pospelov and A.~Ritz,
  Annals Phys.\  {\bf 318} (2005) 119
  [hep-ph/0504231].

\bibitem{Raven:2012fb}
  G.~Raven [LHCb Collaboration],
  arXiv:1212.4140 [hep-ex].


\bibitem{Chatrchyan:2013bka} 
  S.~Chatrchyan {\it et al.} [CMS Collaboration],
  Phys.\ Rev.\ Lett.\  {\bf 111}, 101804 (2013)
  doi:10.1103/PhysRevLett.111.101804
  [arXiv:1307.5025 [hep-ex]].


\bibitem{Buras:2013uqa}
  A.~J.~Buras, R.~Fleischer, J.~Girrbach and R.~Knegjens,
  JHEP {\bf 1307} (2013) 77
  doi:10.1007/JHEP07(2013)077
  [arXiv:1303.3820 [hep-ph]].

\bibitem{Bobeth:2013uxa} 
  C.~Bobeth, M.~Gorbahn, T.~Hermann, M.~Misiak, E.~Stamou and M.~Steinhauser,
  Phys.\ Rev.\ Lett.\  {\bf 112}, 101801 (2014)
  doi:10.1103/PhysRevLett.112.101801
  [arXiv:1311.0903 [hep-ph]].


\bibitem{Becirevic:2012fy} 
  D.~Becirevic, N.~Kosnik, F.~Mescia and E.~Schneider,
  Phys.\ Rev.\ D {\bf 86}, 034034 (2012)
  doi:10.1103/PhysRevD.86.034034
  [arXiv:1205.5811 [hep-ph]].
  
\end{thebibliography}
\end{document}